\documentclass[twocolumn,aps,prb,superscriptaddress,showpacs,floatfix]{revtex4-1}
\usepackage{epsf}
\usepackage{graphicx}
\usepackage{epstopdf}
\usepackage{tikz}
\usepackage{float}
\usepackage{amsmath}
\usepackage{amssymb}
\usepackage{bm}
\usepackage[normalem]{ulem}
\usepackage{soul}

\newcommand{\mb}[1]{ { \mbox{\boldmath{$#1$}}}  }

\begin{document}

\title{Majorana quasiparticles of inhomogeneous Rashba chain}
 
     \author{Maciej M. Ma\'ska}
     \email{maciej.maska@phys.us.edu.pl}
     \author{Anna Gorczyca-Goraj}
     \affiliation{Department of Theoretical Physics, University of Silesia, Katowice, Poland}
     \author{Jakub Tworzyd\l o}
     \affiliation{Faculty of Physics, Warsaw University, Warsaw, Poland}
     \author{Tadeusz Doma\'nski}
 \email{doman@kft.umcs.lublin.pl}
    \affiliation{Institute of Physics, M. Curie-Sk{\l}odowska University, Lublin, Poland}

\date{\today}

\begin{abstract}
We investigate the inhomogeneous Rashba chain coupled to a superconducting substrate,  
hosting the Majorana quasiparticles  near its edges. We discuss its subgap spectrum 
and study how robust are the zero-energy quasiparticles against the diagonal and 
off-diagonal disorder. Studying the $\mathbb{Z}_2$ topological invariant 
we show that disorder induced transition from the topologically non-trivial to 
trivial phases is manifested by characteristic features in the spatially-resolved 
quasiparticle spectrum at zero energy. We provide  evidence for the non-local 
nature of the zero-energy Majorana quasiparticles, that are well preserved upon
partitioning the chain into separate pieces. Even though the Majorana quasiparticles 
are not completely immune to inhomogeneity we show that they can spread onto other 
(normal) nanoscopic objects via the proximity effect.
\end{abstract}


\maketitle

\section{Introduction}
Quasiparticles induced at the edges of spinless ($p$-wave) superconducting 
samples in 1 or 2 dimensions have the exotic character of zero-energy 
bound states \cite{Kitaev-2001,Read-2000,Volovik-1999}. These emergent 
Majorana-type objects have been predicted in various systems, such as
topological insulators \cite{Fu-2009,Tanaka-2009}, semiconducting 
nanowires \cite{Oreg-2010,Lutchyn-2010}, ferromagnetic chains coupled 
to $s$-wave superconducting reservoirs \cite{Choy-2011} etc. Their possible
realizations have been also considered  in topological superconductors 
with electrostatic defects \cite{Tworzydlo-2010}, Josephson type junctions 
\cite{Aguado-2012}, quantum dot chains \cite{Sau-2012,Fulga-2013,Dai-2015},
noncentrosymmetric superconductors \cite{Sato-2009}, ultracold atom systems 
\cite{ultracold} and many other. Intensive studies 
of the Majorana quasiparticles have been overviewed by several authors 
\cite{Alicea-12,Flensberg-12,Stanescu-13,Beenakker-13,Franz-15,Sato-16}. 

The most convincing experimental evidence for the zero-energy Majorana 
modes has been provided so far by the tunneling measurements using the nanoscopic 
chains proximity-coupled to the s-wave superconducting reservoirs 
\cite{Mourik-12,Yazdani-14,Kisiel-15,Franke-15}.
The Majorana quasiparticles are driven at the edges of such chains by 
the strong spin--orbit coupling in presence of the Zeeman splitting, 
when the induced pairing evolves into the topologically nontrivial $p$-wave 
superconductivity of identical spin electrons on the neighboring sites 
\cite{Oreg-2010,Lutchyn-2010}. 

Empirical signatures of the zero-energy quasiparticles have been seen
in the subgap spectroscopy. The first indication was an enhancement 
of the zero-bias differential conductance of the tunneling current 
flowing through the end states of InSb nanowire placed between the conducting 
(Au) electrode and the classical (Nb) superconductor \cite{Mourik-12}. 
The same effect has been later on reported in STM-type configuration, 
by measuring the spatially-resolved differential conductance of magnetic 
(Fe) atom chain deposited on the superconducting (Pb) substrate 
\cite{Yazdani-14,Kisiel-15}. Similar STM setup has been recently 
employed using the superconducting tip \cite{Franke-15}.
Another evidence for the Majorana modes has been reported in 
Bi$_{2}$Te$_{3}$/NbSe$_{2}$ heterostructure by means of the spin-resolved 
Andreev spectroscopy \cite{Sun-2016}.

%
\begin{figure} 
\includegraphics[width=0.9\linewidth]{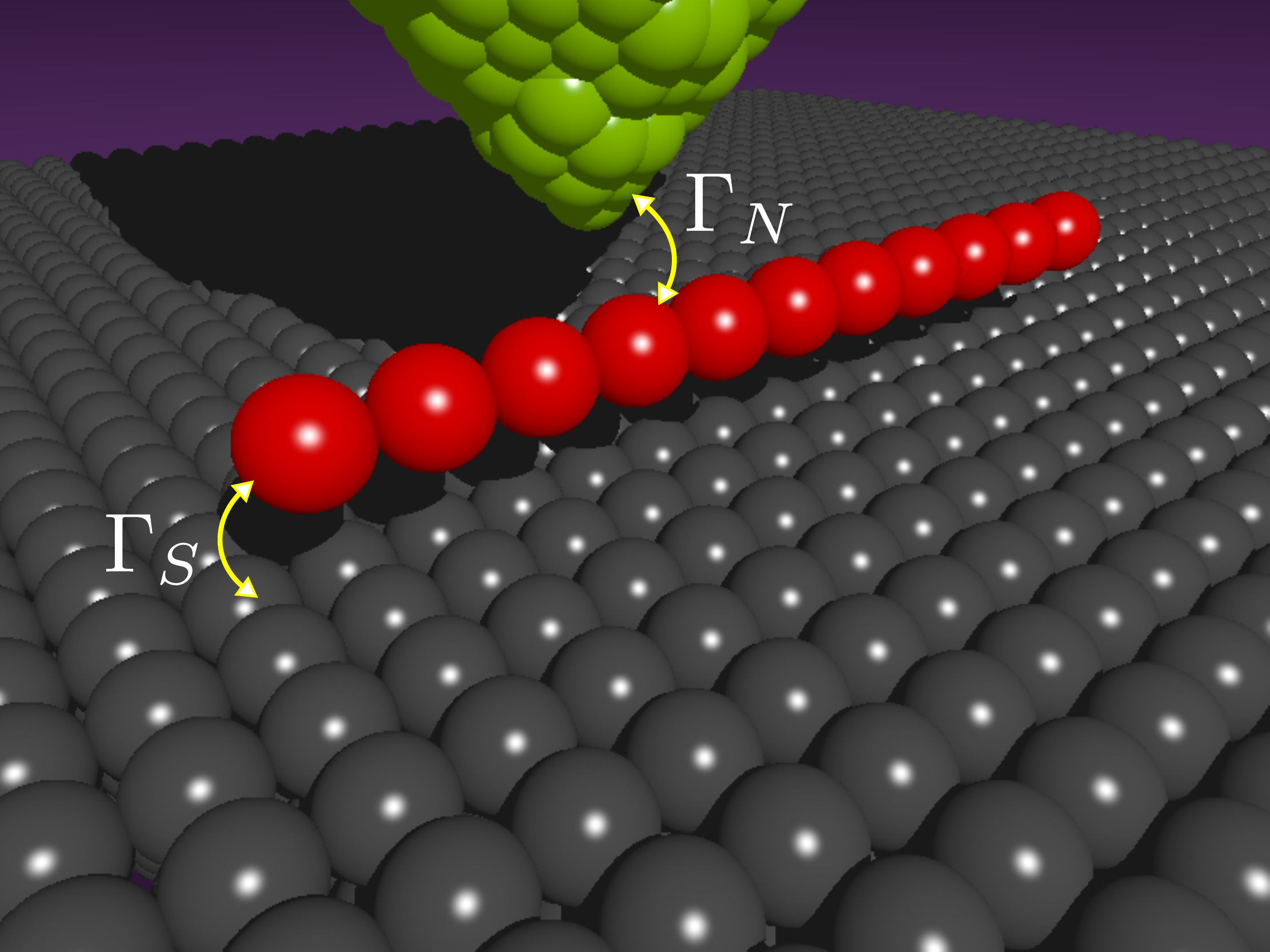}
\caption{Schematic view of the STM setup, in which the atomic chain 
(hosting the Majorana quasiparticles) is deposited on  s-wave 
superconducting substrate and is probed by the charge current of 
the normal (conducting) tip.  }
\label{fig1}
\end{figure}
%

The purpose of our work is to study a stability of the zero-energy Majorana modes 
in STM-type configuration (Fig.\ \ref{fig1}), relevant to the  experiments 
of Princeton  \cite{Yazdani-14} and Basel \cite{Kisiel-15} groups. In realistic 
situations the Rashba chain on a surface of s-wave superconductor could be 
affected by inhomogeneity of: (i) atom energies (diagonal disorder), 
(ii) coupling to the superconducting substrate (off-diagonal disorder), 
and (iii) intersite hopping integral. It is hence important to study 
how robust are the Majorana modes to various disorders. Some aspects 
of the inhomogeneous Rashba chains have been already addressed by several 
groups \cite{Stanescu-2011,Brouwer-2011a,Brouwer-2011b,Rainis-2013,
DeGottardi-2013,Hu-2015,Hui-2015,Pekerten-2015,Klinovaja-2015,
Klinovaja-2016,Nori-16,Wang-2016,Cole-2016,Hegde-2016},
emphasizing that the Majorana quasiparticles are not completely immune 
to the moderate and strong disorder \cite{Nori-16}. Here we revisit this 
issue in a systematic way. In particular, we ($i$) study the 
boundary effects and their influence on spatial extent of the 
zero-energy bound states, ($ii$) consider disordered- induced 
transition from the topologically non-trivial to trivial states, 
($iii$) analyze stability of the Majorana quasiparticles upon partitioning 
the chain into pieces, and ($iv$) present how  the Majorana quasiparticles
spread on other side-attached nanoobjects (quantum impurities) 
via the proximity effect. 

The paper is organized as follows. In section II we  formulate the 
microscopic model and  discuss the subgap spectrum of a finite-length 
Rashba chain. In section III we consider stability of the Majorana states 
in the inhomogeneous chain against the random atom energies and coupling 
to the superconducting substrate. Next, in section IV, we consider an 
interplay between the single impurities and the Majorana states. 
Finally, in section V, we summarize our results  and present some 
technical details in Appendices A-D.

\section{Microscopic model}

STM-type configuration \cite{Yazdani-14,Kisiel-15} (displayed in 
Fig.\ \ref{fig1}) can be modeled by the following Hamiltonian
\begin{eqnarray} 
\hat{H} &=& \hat{H}_{\rm tip} + \hat{H}_{\rm chain} + \hat{H}_{S} 
+  \hat{V}_{\rm hybr}  ,
\label{model}
\end{eqnarray} 
where $\hat{H}_{\rm tip}$ describes the normal tip, $\hat{H}_{\rm chain}$ refers to 
the atomic chain, and $\hat{H}_{S}$ stands for the s-wave superconducting 
substrate. For specific considerations we describe the atomic chain by 
the tight-binding model 
$\hat{H}_{\rm chain}=\sum_{i,\sigma}   \varepsilon_{i} 
\hat{d}^{\dagger}_{i,\sigma} \hat{d}_{i,\sigma}+\sum_{i,j,\sigma}
t_{ij}\hat{d}^{\dagger}_{i,\sigma} \hat{d}_{j,\sigma} + 
\hat{H}_{\rm Rashba} + \hat{H}_{\rm Zeeman}$, where the second quantization operator 
$\hat{d}_{i,\sigma}^{(\dag)}$ annihilates (creates) an electron 
 at $i$-th site with energy $\varepsilon_{i}$ and spin 
$\sigma$, and the inter-site hopping integral is denoted by $t_{ij}$. 
We assume the magnetic field {\bf B}$=(0,0,B)$ and 
the spin--orbit vector $\bm{\alpha}=(0,0,\alpha)$. We  express the Rashba  
interaction and the Zeeman terms by
\begin{eqnarray}
    \hat{H}_\mathrm{Rashba}&=&-\alpha\sum_{i,\sigma,\sigma'}\left[
    \hat{d}^{\dagger}_{i+1,\sigma}
    \left(i\sigma^y\right)_{\sigma\sigma'}\hat{d}_{i,\sigma'}+\mathrm{h.c.}\right],\\
    \hat{H}_\mathrm{Zeeman}&=&\frac{g\mu_\mathrm{B}B}{2}\sum_{i,\sigma,\sigma'}
    \hat{d}^{\dagger}_{i,\sigma}\left(\sigma^z\right)_{\sigma\sigma'}
    \hat{d}_{i,\sigma'}.
\end{eqnarray}

We treat the STM tip as the free fermion gas 
$\hat{H}_{N} \!=\! \sum_{{\bf k},\sigma} \xi_{{\bf k}N}  
\hat{c}_{{\bf k} \sigma N}^{\dagger} \hat{c}_{{\bf k} \sigma N}$   
and describe the isotropic superconductor by the BCS model
$\hat{H}_{S} \!=\!\sum_{{\bf k},\sigma}  \xi_{{\bf k}S}
\hat{c}_{{\bf k} \sigma S }^{\dagger}  \hat{c}_{{\bf k} \sigma S} 
\!-\! \sum_{\bf k} \Delta_{sc}  \left( \hat{c}_{{\bf k} \uparrow S }
^{\dagger} \hat{c}_{-{\bf k} \downarrow S }^{\dagger} + \hat{c}
_{-{\bf k} \downarrow S} \hat{c}_{{\bf k} \uparrow S }\right)$.
The operators $\hat{c}_{\mathbf{k}\sigma\beta}^{(\dag)}$ refer 
to the itinerant electrons with momentum $\mathbf{k}$, spin 
$\sigma$, and energy $\xi_{{\bf k}\beta}=\varepsilon_{\bf k}
-\mu_{\beta}$  (where $\beta=N,S$). 
Hybridization between the atoms and both external 
reservoirs is described $\hat{V}_{\rm hybr} 
= \sum_{{\bf k},\sigma,\beta}  \left( V_{i,{\bf k} \beta} \; 
\hat{d}_{i,\sigma}^{\dagger}  \hat{c}_{{\bf k} \sigma \beta } 
+  V_{i,{\bf k} \beta}^{*}  \hat{c}_{{\bf k} \sigma 
\beta }^{\dagger} \hat{d}_{i,\sigma} \right)$, where 
$V_{i,{\bf k} \beta}$ are the tunneling matrix elements.

\subsection{Deep subgap regime}

Since the zero-energy modes are formed inside the superconducting
energy regime $(-\Delta_{sc},\Delta_{sc})$ it is convenient to introduce 
the characteristic couplings $\Gamma_{i,\beta}=2\pi \sum_{\bf k} 
|V_{i,{\bf k}\beta}|^2 \; \delta(\omega - \xi_{{\bf k}\beta})$ 
and treat them as constant quantities. In the weak coupling limit, 
$\Gamma_{S} \ll \Delta$,  the superconducting reservoir induces
the electron pairing at each of the atoms (see Appendix A for details)
\begin{eqnarray} 
\hat{H}_{S} + \sum_{{\bf k},\sigma}  \left( V_{i,{\bf k} S} \; 
\hat{d}_{i,\sigma}^{\dagger}  \hat{c}_{{\bf k} \sigma S} 
+  \mbox{\rm h.c.} \right)  
\equiv 
\Delta_{i} \hat{d}_{i,\uparrow}^{\dagger} 
\hat{d}_{i,\downarrow}^{\dagger} + \mbox{\rm h.c.} 
\label{large_Delta} 
\end{eqnarray} 
The proximity-induced pairing potential is $\Delta_{i}=\Gamma_{i,S}/2$.
Atomic chain can be hence formally described by the following
low-energy Hamiltonian \cite{Stanescu-13}
\begin{eqnarray} 
\hat{H}_{\rm chain}^{\rm (prox)} &=&
  \sum_{i,\sigma}  \varepsilon_{i} \hat{d}^{\dagger}_{i,\sigma} 
\hat{d}_{i,\sigma} + \sum_{i,j,\sigma}t_{ij}\hat{d}^{\dagger}_{i,\sigma} 
\hat{d}_{j,\sigma}  + \hat{H}_{\rm Rashba} 
\nonumber \\ & + &  \hat{H}_{\rm Zeeman} + \sum_{i} \Delta_{i} \left( 
\hat{d}_{i,\uparrow}^{\dagger} \hat{d}_{i,\downarrow}^{\dagger} 
+ \hat{d}_{i,\downarrow} \hat{d}_{i,\uparrow}  \right). 
\label{proximized}
\end{eqnarray} 
In a homogeneous system
($\varepsilon_i=\mu,\ t_{ij}=t\delta_{|i-j|,1}$,  $\Delta_i=\Delta$) 
the zero-energy quasiparticles of (\ref{proximized}) exist in the
topologically non-trivial superconducting state, in a region 
restricted by the boundaries \cite{Oreg-2010,Lutchyn-2010,Gibertini}
\begin{equation}
  \left(\mu \pm 2t\right)^2+\Delta^2-V_{\rm Z}^2=0,
  \label{eq:boundaries}
\end{equation}
where $V_{\rm Z}=g\mu_{\rm B}B/2$ is the Zeeman energy. Majorana-type 
quasiparticles of the inhomogeneous systems (for various kinds of 
disorder) are discussed in Sec.\ III.

Obviously this heuristic scenario (\ref{proximized}) does not capture any
electronic states existing outside the energy gap of bulk superconductor
$|\omega| > \Delta_{sc}$. To take them into account one should properly treat
the dynamic effects \cite{Stanescu-2011,Peng-2015} appearing in the energy-dependent
selfenergy (mentioned in Appendix A). Other possibility would be to study the
Rashba chain along with the piece of superconducting substrate by the Bogoliubov--de
Gennes approach \cite{Chevallier-2012}. Such states would eventually induce
a continuous background of the high-energy spectrum, and could show up in the
tunneling characteristics at $e|V|\geq \Delta_{sc}$.  The Rashba chain weakly coupled 
to superconducting substrate the subgap states is at low energies reliably 
reproduced by the static approximation  (\ref{proximized}) \cite{Stanescu-13}. 
Our present study is focused here on stability of the zero-energy quasiparticles, 
therefore we skip the high-energy effects.

\subsection{Intrinsic inhomogeneity of atomic chain\label{fse}}

Let us briefly analyze the in-gap quasiparticles of the uniform chain 
consisting of $N$ atoms. We have determined numerically the eigenvalues 
and eigenvectors of the proximized atomic chain (\ref{proximized}) for 
$N=70$, using the following model parameters: $\varepsilon_{i}/t=-2.1$, 
$\alpha/t=0.15$ and  $\Gamma_{i,S}/t=0.2$. These parameters 
are chosen to guarantee that the system is in topologically nontrivial
regime, unless the critical disorder is achieved.
%
\begin{figure} 
\includegraphics[width=0.85\linewidth]{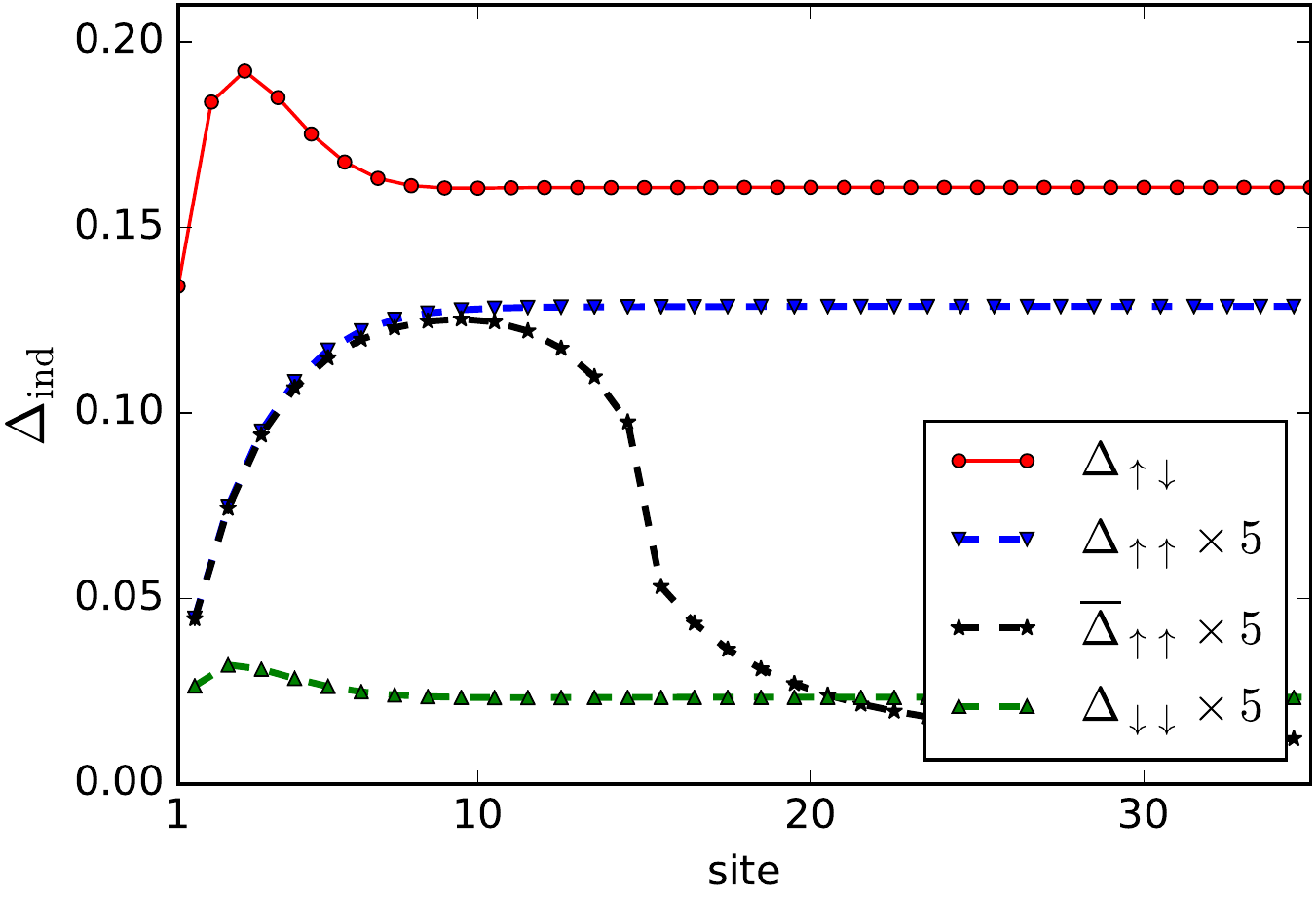}
\caption{The spatially resolved induced pairings
  $\Delta_{\uparrow\downarrow}=\langle \hat{d}_{i,\downarrow} \hat{d}_{i,\uparrow}\rangle$,
  $\Delta_{\uparrow\uparrow}=\langle \hat{d}_{i,\uparrow} \hat{d}_{i+1,\uparrow}\rangle$, and
  $\Delta_{\downarrow\downarrow}=\langle \hat{d}_{i,\downarrow} \hat{d}_{i+1,\downarrow}\rangle$
obtained for the same model parameters as in Fig.\ \ref{fig2}
and $g\mu_{B}B/2=0.27t$. $\overline{\Delta}_{\uparrow\uparrow}$ denotes the magnitude
of the order parameter $\langle \hat{d}_{i,\uparrow} \hat{d}_{i+1,\uparrow}\rangle$
calculated in the case when the spin--orbit interaction is non--zero only close to
the ends of the chain (on sites from 1 to 15 and from 56 to 70). For the sake of visibility
the intersite equal-spin pairing amplitudes are multiplied by~5.}
\label{fig3}
\end{figure}
%

In the popular Kitaev model the Majorana bound states appear at the very 
last sites of one-dimensional chain characterized by the uniform intersite 
triplet pairing. In reality, however, a magnitude of the induced $p$-wave 
pairing would be affected by the finite atomic length. To get some insight 
into such effects we investigate the pairing amplitudes of both the singlet 
$\langle \hat{d}_{i,\downarrow} \hat{d}_{i,\uparrow} \rangle$ and 
equal--spin channels $\langle \hat{d}_{i,\sigma} \hat{d}_{i+1,\sigma} \rangle$, 
respectively. Fig.~\ref{fig3} shows their spatial variation for 
the chosen model parameters. The internal chain sites are characterized 
by nearly constant (uniform) value of the pairing amplitude, whereas at 
the edges there appear some deviations. The maximal amplitude of $\langle 
\hat{d}_{i,\downarrow} \hat{d}_{i,\uparrow} \rangle$ corresponds to
$i \sim 3$ (and $i \sim N-3$). The other peripheral chain atoms 
$i=1,2$ are characterized by the clearly reduced pairing amplitude. 
This intrinsic inhomogeneity has noticeable implications 
on the induced $p$--wave pairing and such aspect distinguishes our approach 
from the Kitaev toy model \cite{Kitaev-2001}. Practically, such 
effects could be verified by means of the spin-polarized Andreev spectroscopy  
\cite{Sun-2016}.

\begin{figure} 
    \includegraphics[width=0.85\linewidth]{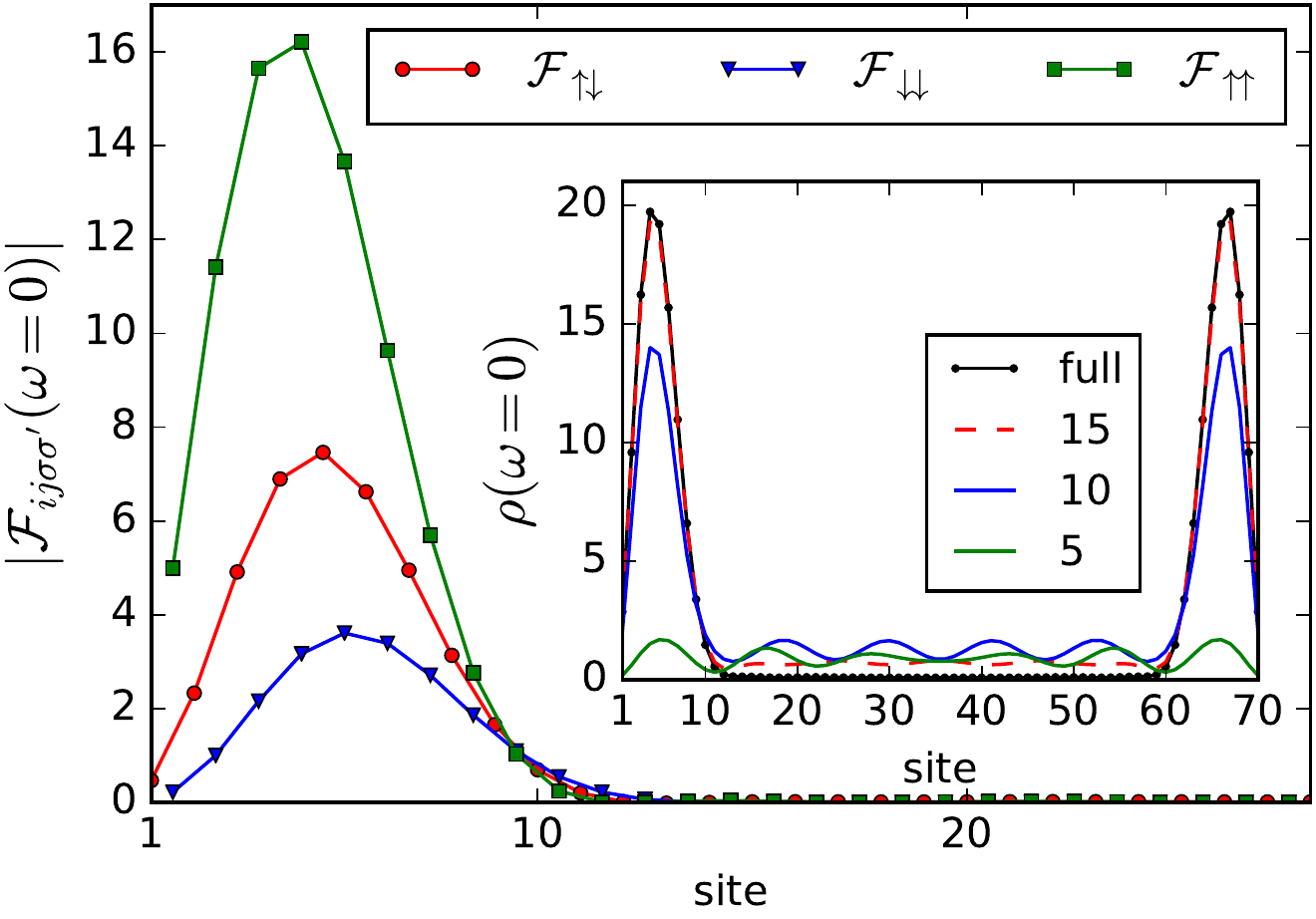}
    \caption{Off--diagonal spectral densities ${\cal F}_{ij\sigma\sigma^\prime}(\omega)$ at $\omega=0$ for different pairings.
      The inset shows a comparison of the zero--energy local density of states
      for different sizes of the part of the chain where the spin--orbit
      interaction is switched off. The solid black line (``full'') represents the reference
      chain with the spin--orbit interaction on all lattice sites. Lines described
      as ``15'', ``10'', and ``5'' show the LDOS in cases when the interaction is present
      only on 15, 10, and 5 outermost sites, respectively. Note, that the line marked
      as ``15'' differs from the reference line only in the central region.}
    \label{spectral_densities} 
  \end{figure}

Spatial profile of the Majorana quasiparticles 
is essentially dependent on the anomalous  spectral densities 
${\cal  F}_{ij\sigma\sigma^\prime}(\omega)=-\frac{1}{\pi}{\rm Im}\langle\langle
\hat{d}_{i,\sigma}; \hat{d}_{j,\sigma^\prime} \rangle\rangle$ at zero energy.  Fig.~\ref{spectral_densities} shows them for the $s$-wave
(where $j=i$ and $\sigma\ne\sigma^\prime$) and the $p$-wave
(where $j=i+1$ and $\sigma=\sigma^\prime$) pairing channels. 
In both cases the anomalous spectral function 
${\cal  F}_{ij\sigma\sigma^\prime}(\omega\!=\!0)$
does not vanish only in such regions where the Majorana 
states exist. The induced amplitudes (for each  pairing channel 
shown in Fig.\ \ref{fig3}) have been calculated from 
$\langle \hat{d}_{i,\sigma} \hat{d}_{j,\sigma^\prime}\rangle
=\int d\omega\, {\cal  F}_{ij\sigma\sigma^\prime}(\omega) \,
f(\omega,T)$, where $f(\omega,T)=\left[1+\mbox{\rm exp}(\omega/k_{B}T)
\right]^{-1}$ is the Fermi-Dirac distribution function.

%
\begin{figure} 
\includegraphics[width=0.995\linewidth]{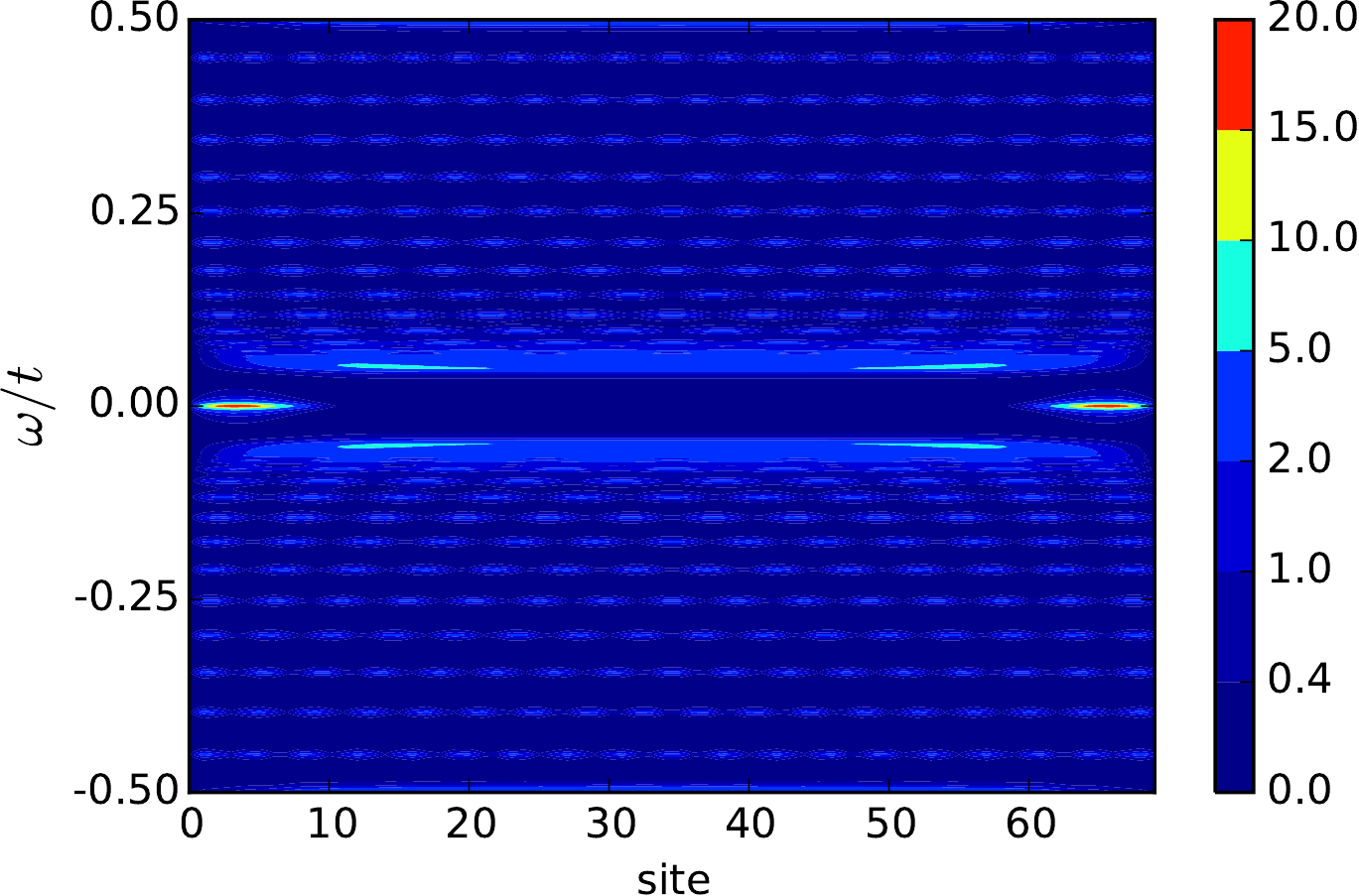}
\caption{Subgap spectrum of the atomic chain 
for each site $i\in \left< 1, 70 \right>$ 
with the line-broadening imposed by $\Gamma_{N}=0.1\Gamma_{S}$.}
\label{fig4}
\end{figure}
%

\begin{figure*}[htb]
\includegraphics[width=0.99\columnwidth]{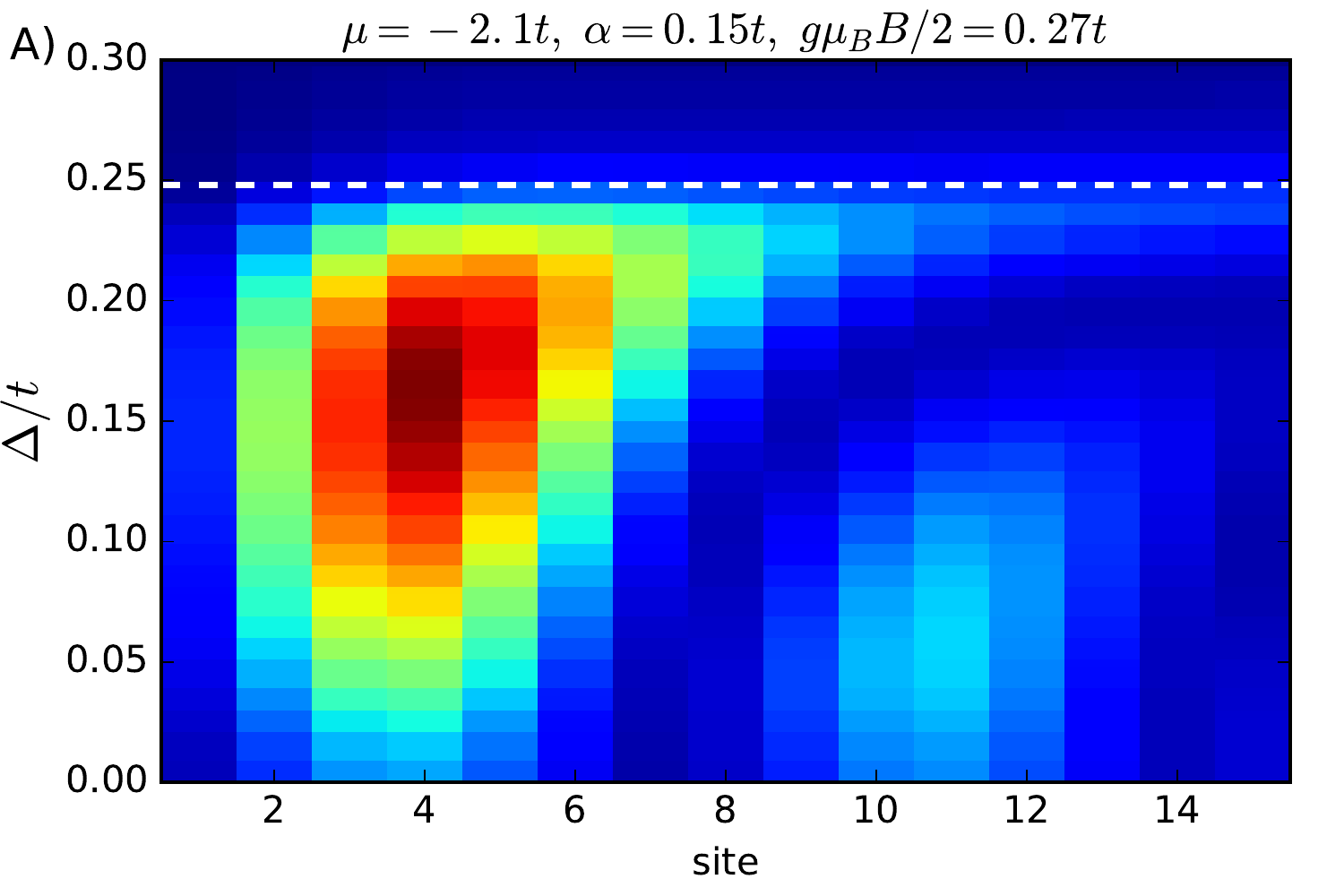}
\includegraphics[width=0.99\columnwidth]{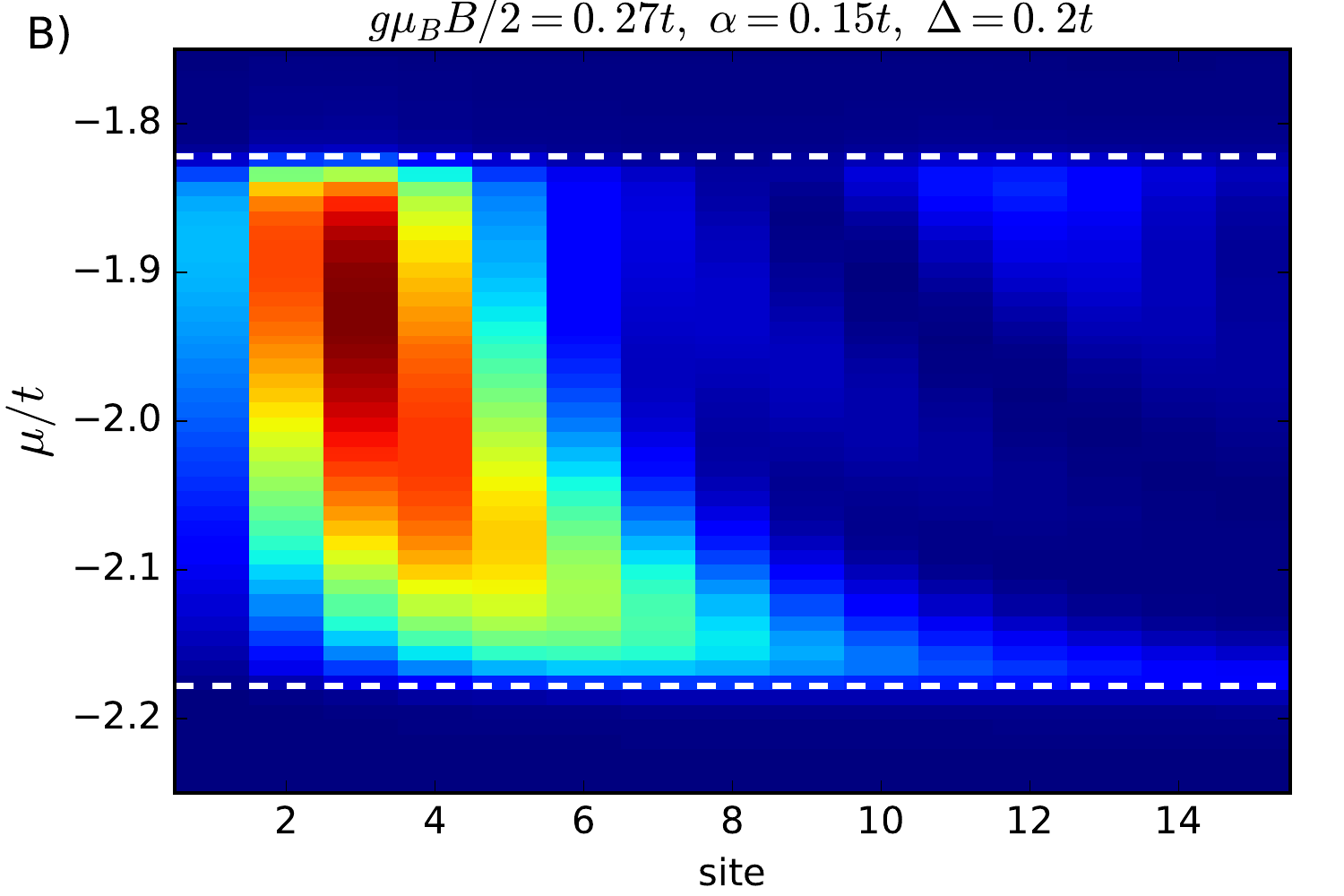}
\includegraphics[width=0.99\columnwidth]{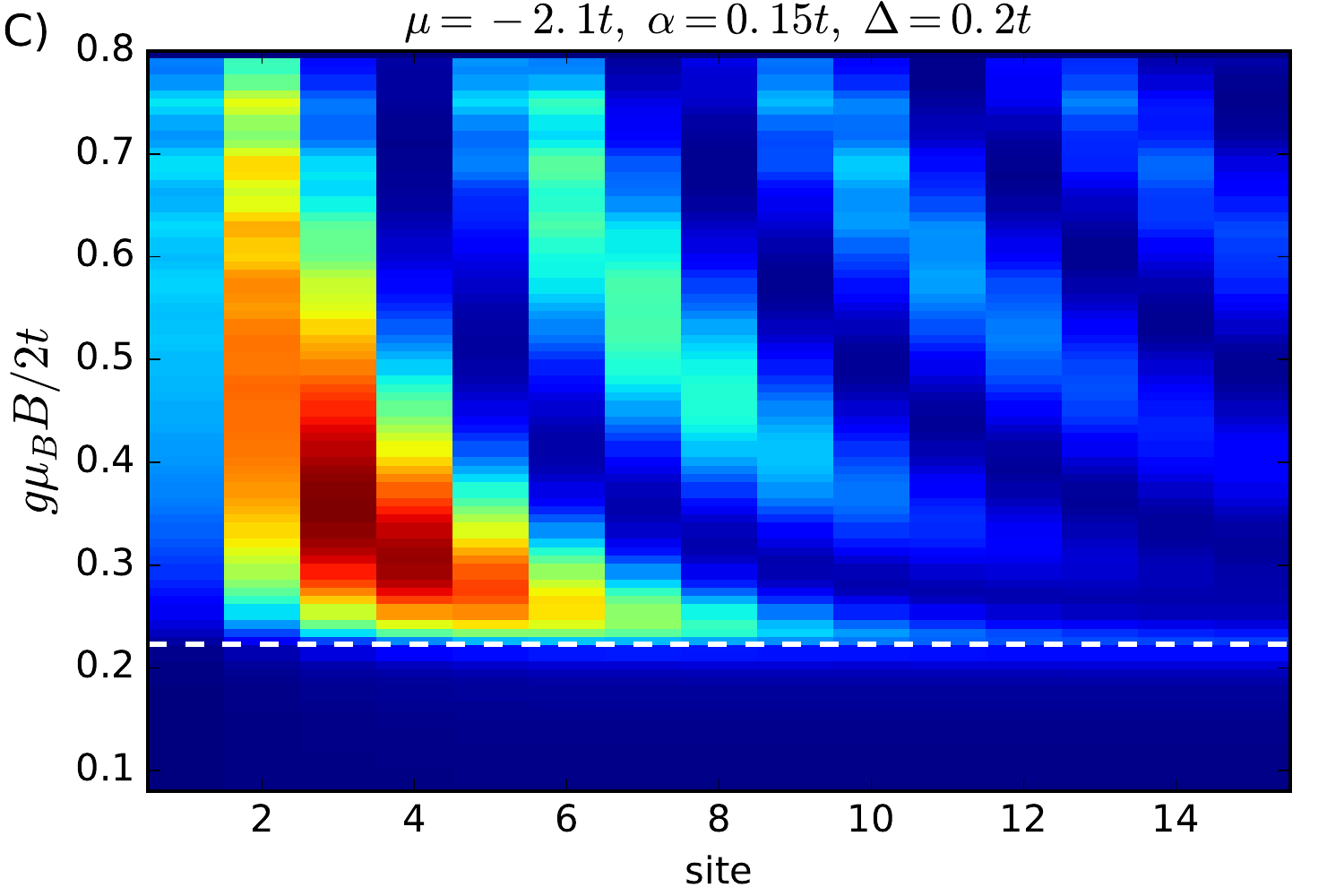}
\includegraphics[width=0.99\columnwidth]{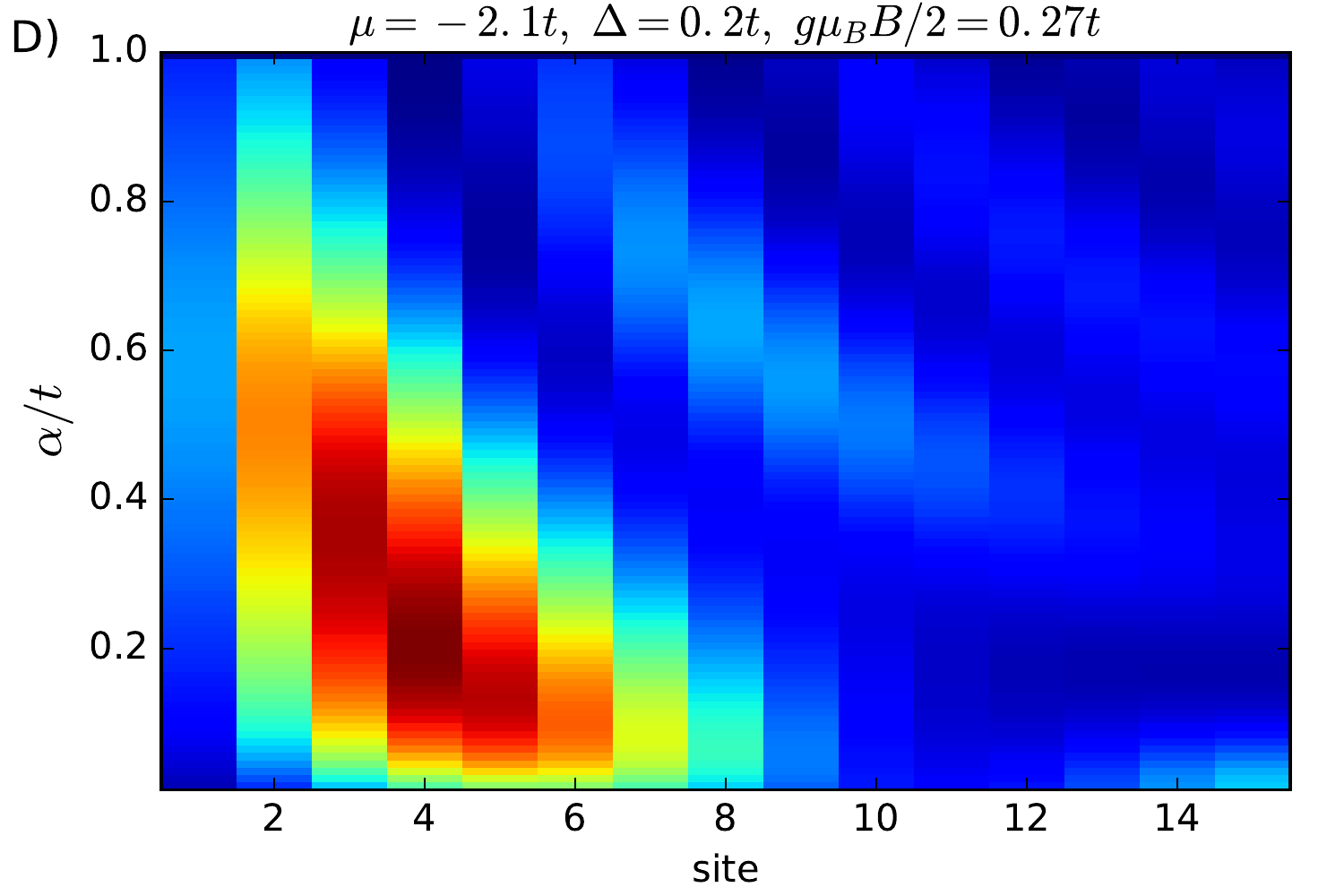}
\caption{Local DOS at $\omega=0$ as a function of the model parameters:
  A)~magnitude of the induced pairing $\Delta$, 
  B)~the chemical potential
  $\mu$, C)~magnetic field $B$, and D)~the spin--orbit coupling $\alpha$.
  The values of the other (fixed) parameters are given above each panel.
  The white dashed lines show boundaries of the topological region given by
  Eq.~(\ref{eq:boundaries}). Note, that independently of the model parameters
  the LDOS maximum is always shifted from the nanowire edge.
\label{fig:M_params}}
\end{figure*}

Since the spectral functions ${\cal  F}_{ij\sigma\sigma^\prime}(\omega=0)$
vanish away from the Majorana states, we presume that electron pairing 
would be necessary only at the chain edges. We checked such a possibility 
by switching off the spin--orbit interaction (responsible for $p$-wave 
pairing) on internal sites of the chain. Our results (see inset in
Fig.~\ref{spectral_densities}) prove, that  as long as the region 
of the absent spin--orbit interaction does not coincide with the 
Majorana quasiparticles, their profile is not really much affected. 
For instance, the red dashed line in the inset in
Fig. \ref{spectral_densities} (corresponding to the spin--orbit interaction
present only in 15 sites 
at the chain edges) is nearly identical with the result for uniform 
system. By further expanding the region of absent spin-orbit interaction 
the Majorana states become eventually damped. Simultaneously we 
observe small oscillations of $\rho(\omega=0)$  appearing in part 
of the atomic chain where the spin--orbit coupling is absent. This is 
consistent with the results reported in Ref. \onlinecite{Wang-2016},
for the Rashba chain with non-uniform spin--orbit coupling. By switching off 
the spin--orbit coupling in a half of the chain, the Majorana quasiparticle 
of the ``noninteracting'' half evolved into the finite-energy 
Shiba/Andreev states. Our system can be regarded as two pieces of 
such chains, interconnected by the ``noninteracting'' parts with 
two Majorana quasiparticles preserved at the opposite edges. Another 
relative situation will be discussed in Sec. \ref{sec:partitioning}, 
where we consider a gradual partitioning of the atomic chain.

The boundary effects  have also influence on a profile 
of the Majorana quasiparticles \cite{Alicea-16,Klinovaja-2016-topography}. 
In Fig.\ \ref{fig4} we present the spatially resolved local density of 
states (LDOS). The Majorana quasiparticles spread over nearly 10 peripheral 
atoms and (for the chosen model parameters) their maximal 
intensity occurs at sites $i=4$ and $i=66$. 
This fact nicely coincides with the real experimental data, reported 
by the Princeton group \cite{Yazdani-14}. The strongest zero-bias 
enhancement of the subgap STM tunneling corresponds to the point 
'2' in Fig. 4 of the report by E.~Yazdani  \cite{Yazdani-2015}. 
Its distance from the chain edge is roughly $\sim$ 8 \AA, 
so the Majorana feature is indeed centered near 
the 4$^{\mbox{\footnotesize \rm th}}$ iron atom.

We have checked, that such maximum of the local 
density of states rather weakly depends on the model parameters 
[provided that the system is in the topologically nontrivial 
regime given by Eq.\ (\ref{eq:boundaries})].  Spatial profile 
of the zero-energy Majorana quasiparticles and their dependence
on $\Delta$, $\mu$, $B$, and $\alpha$ is presented in Fig. 
\ref{fig:M_params}. We clearly notice that the maximal intensity 
roughly appears either at the 4$^{\mbox{\footnotesize \rm th}}$ or 
3$^{\mbox{\footnotesize \rm rd}}$ atom from the chain edge.

LDOS at zero energy is also sensitive to the localization length
of the Majorana quasiparticles and the interference between 
different components of the ``composite'' wave function
\cite{Oscillations1,Oscillations2,Oscillations3,Klinovaja-2012},
\begin{equation}
\Psi^M(x)\sim \sin(k_Fx)e^{-x/\xi} , 
\end{equation}
where $k_F$ depends on microscopic parameters, such as the chemical 
potential or the Zeeman splitting. The localization length $\xi$ itself, 
however, is not sufficient to explain the position of the LDOS maximum.
For example, the actual maximum slightly departs from the chain edge 
upon increasing $\Delta$  (see Fig. \ref{fig:M_params}A), whereas 
the localization length is expected to scale in opposite way 
$\xi \propto 1/\Delta$ \cite{Peng-2015}.
This suggests that the boundary effects within the low-energy 
effective model (\ref{proximized}) describe the realistic situation, 
without any need for fine--tuning to reproduce the experimentally 
observed shape of the Majorana bound states. 

\section{Disordered Rashba chain}

In this section we investigate whether the Majorana quasiparticles can
sustain such forms of the inhomogeneity as the random site energies 
$\varepsilon_{i}$ (we call it `diagonal disorder') and the spatially varying 
coupling $\Gamma_{i,S}$ to s-wave superconducting substrate that affects 
the pairing potential $\Delta_{i}$ (we call it `off-diagonal disorder'). 

\subsection{Random site energies}

\begin{figure} 
\includegraphics[width=0.85\linewidth]{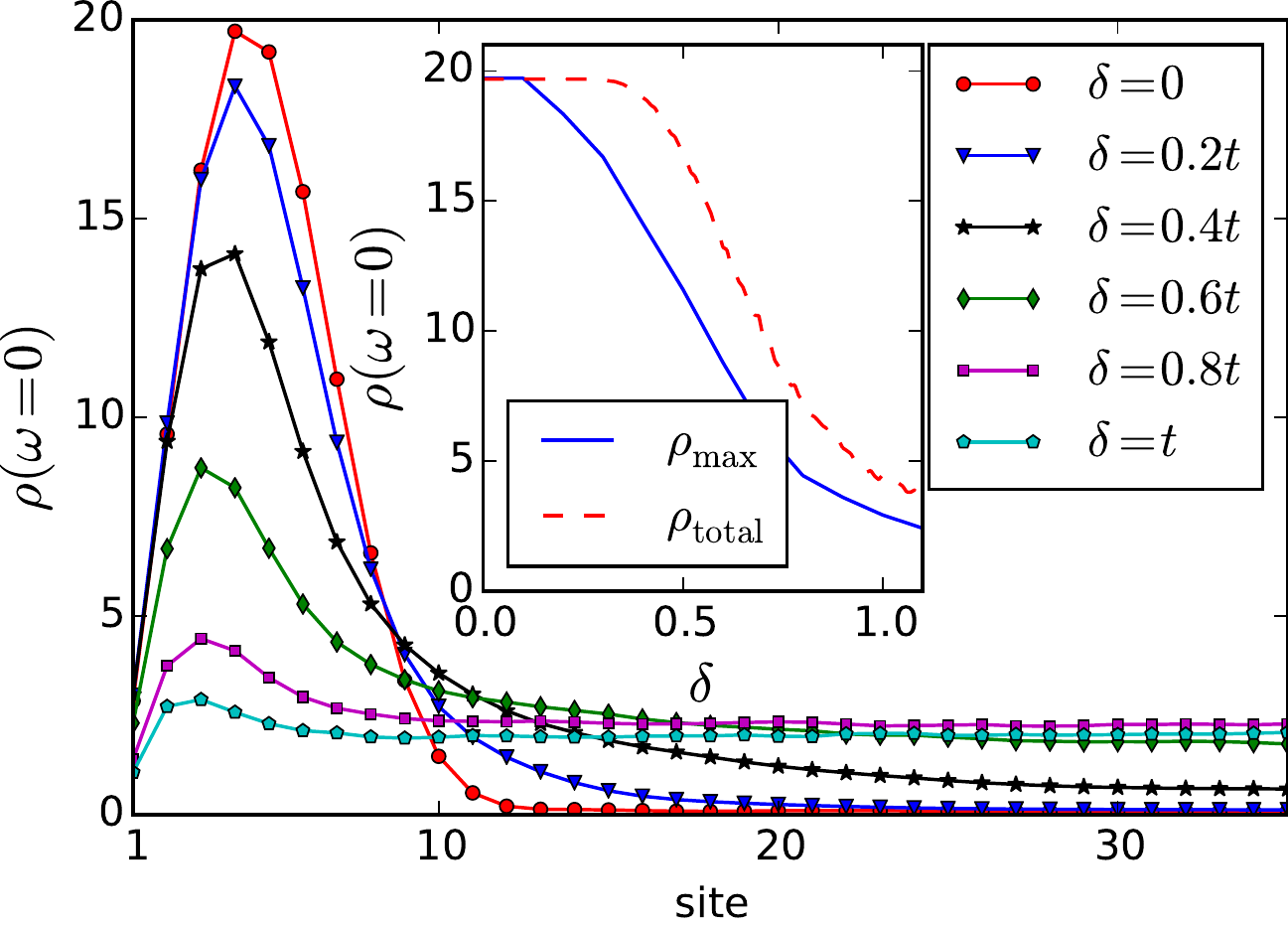}
\caption{Evolution of the zero-energy subgap quasiparticle
spectrum $\rho_{i}(\omega=0)$ driven by the diagonal disorder. 
Inset shows the local density of states obtained at $\omega=0$  
for $i=4$, where the Majorana quasiparticle has the highest 
probability (solid blue line), and the density of states 
averaged over all sites of the chain (rescaled to the same maximum value,
dashed red line).}
\label{diag_disorder}
\end{figure}

We have chosen the energies $\varepsilon_{i}=\varepsilon + \xi_{i}\, \delta$  
with a random number $\xi_{i}\in \left< -1, 1 \right>$ and the magnitude 
$\delta$ ranging from small to large values \cite{Wang-2016}. We have  next
determined the spectral function and averaged it roughly over $\sim 10^{4}$ 
different configurations $\left\{ \xi_{i}\right\}$, depending on $\delta$. 
The main panel in Fig.\ \ref{diag_disorder} displays spatial variation of the 
averaged spectral function at zero energy $\omega=0$ for representative values 
of $\delta$, ranging from the weak  to strong disorder.

With increasing $\delta$ the subgap spectrum is gradually filled-in. To illustrate this behavior  we present in inset to Fig.\ \ref{diag_disorder} the spectral function $\rho_{i}(\omega=0)$ at site $i=4$ (solid blue line), where the Majorana quasiparticle has the largest probability. We observe that near some critical amplitude $\delta^{*} \sim 0.7$ our system undergoes a qualitative changeover, above which  $\rho_{4}(\omega=0)$ asymptotically tends to a value, common for the entire atomic chain. 
The dashed red line in inset to Fig. \ref{diag_disorder} shows the density of states averaged over entire chain. We can notice that for a weak disorder
the averaged zero-energy density remains almost intact. It means
that in presence of the weak disorder the bound states survive at $\omega=0$, but they can be shifted from the chain edges. For stronger disorder the total zero-energy density is reduced,  
indicating that the subgap quasiaprticles move to finite energies.

\begin{figure} 
\includegraphics[width=0.85\linewidth]{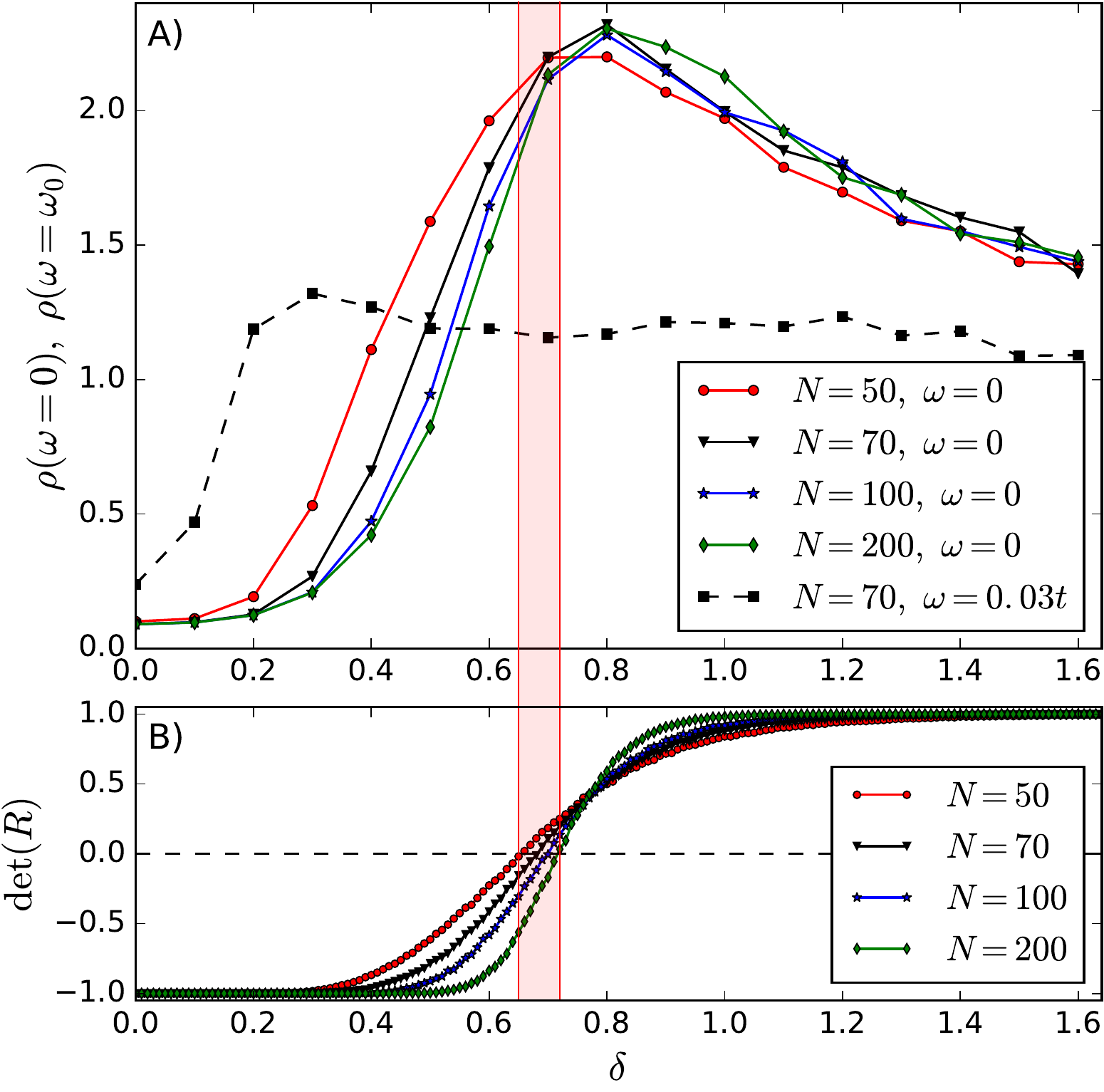}
\caption{The quasiparticle spectrum  $\rho_{i}(\omega)$ at the central site $i=N/2$ averaged over $10^4$ realizations of the random site energies. The solid lines 
in panel A show the density of states at $\omega=0$  for several $N$, as indicated. The black dashed line presents $\rho_{i}(\omega_0)$ at small (yet finite) energy $\omega_0=0.03t$ for $N=70$. 
Panel B presents the determinant of the reflection matrix averaged 
over $10^3$ disorder realizations versus the disorder $\delta$. Sign of this 
determinant changes between the topologically trivial and nontrivial phases, 
as discussed in Sec.\ \ref{topo}. The vertical stripe marks a range of $\delta$ 
in which the averaged $\det(R)$ changes the sign, depending on the chain length 
varying from $N=50$ to $N=200$. We can notice, that the sign change of 
$\det(R)$ nearly coincides with the maximum of $\rho_{i}(\omega=0)$. 
\label{middle_point}}
\end{figure}
%

\begin{figure}
\includegraphics[width=0.85\linewidth]{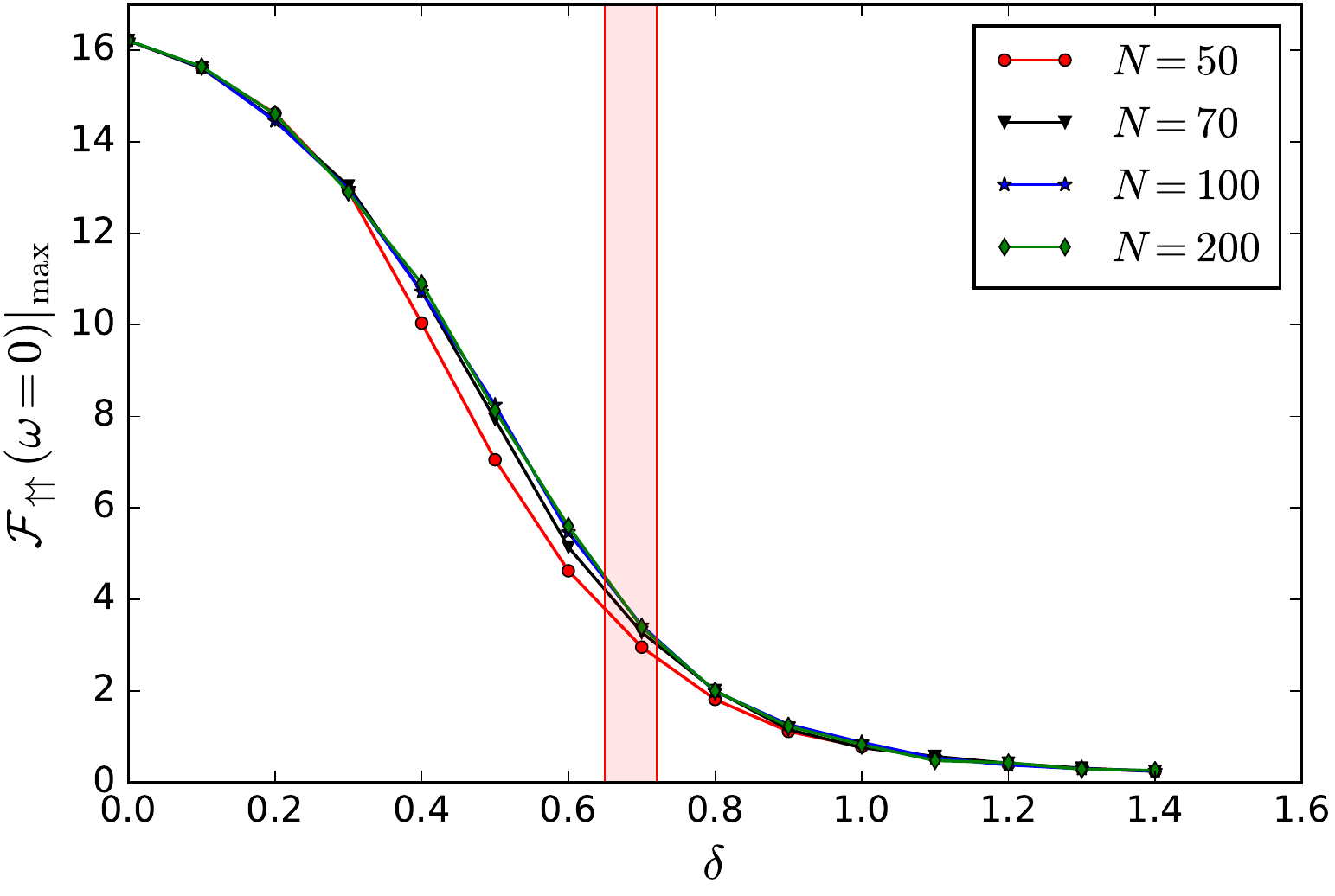}
\caption{Suppression of the $p$-wave superconducting phase induced 
by the disorder, where $\left.{\cal F}_{\uparrow\uparrow}
(\omega=0)\right|_{\rm max}$ is the maximum of the $p$-wave spectral 
density along the chain. The red vertical strip has the same meaning 
as in Fig.~\ref{middle_point}B.}
\label{fig:pwave}
\end{figure}

To further clarify such changeover we show in Fig.\ \ref{middle_point} concomitant evolution of the averaged spectral function $\rho_{i}(\omega=0)$ at the internal site $i=N/2$ (where for $\delta=0$ the zero-energy quasiparticles are absent). With increasing disorder (but below $\delta^{*}$) the zero-energy states gradually build up. We argue that they originate from the Majorana states that are pinned by disorder on internal sites (as explained in section IV). With further increase of $\delta$ (above $\delta^{*}$) the zero energy spectral function slowly diminishes. This tendency is caused by evolution from the nontrivial to trivial superconducting phases. 
Such changeover is smooth, regardless of the atomic chain size  $N$ (see the solid lines in Fig.\ \ref{middle_point}).
Additional indication that strong disorder triggers  the trivial superconducting state is provided by evolution of $\rho_{i}(\omega)$ at finite energy $\omega=0.03t$, corresponding to the soft gap regime (see Appendix B). The black dashed line in Fig.\ \ref{middle_point} indeed shows that strong disorder closes the soft gap, what is evidenced by saturation of $\rho_{i}(\omega=0.03t)$ for large $\delta$, inducing the ordinary Andreev/Shiba states. 
Similar scenario, where the 
Majorana quasiparticles are destroyed by critical disorder holds true 
also for more realistic multiband systems (see Appendix \ref{sec:two-band}).
Additional evidence, that the critical disorder magnitude $\delta^{*}$ 
corresponds to transition to the topologically trivial phase, is
provided in Sec. \ref{topo}.

There is a significant advantage of studying the local density of states
away from the Majorana edge states. As illustrated in Fig. \ref{diag_disorder}, 
the magnitude of the Majorana states smoothly decreases almost to zero with 
increasing disorder without any characteristic points that could be used 
to pinpoint the topological transition. However, in the corresponding
evolution of the zero--energy states away from the chain edges (Fig.\ 
\ref{middle_point}), such a feature is well pronounced and can be 
easily identified.

Fig.\ \ref{fig:pwave} illustrates the zero-energy 
$p$--wave spectral density versus $\delta$ obtained for various $N$.
This result confirms that  above $\delta^{*}$ the Majorana quasiparticles 
disappear as a consequence of the suppressed equal-spin pairing 
and $\delta^{*}$ can be regarded to be characteristic point for 
a crossover to the non-trivial pairing.

What is interesting, for $\delta<\delta^{*}$ the spectral function
$\rho_{i}(\omega=0)$ scales with the system size $N$, whereas only minor
statistical fluctuations show up above $\delta^{*}$. While for any finite
disorder all the states are localized (what is seen by means of the
{\em inverse participation ratio} method, not presented here), 
it may suggest that at $\delta^{*}$ the localization length changes
abruptly. The critical character of $\delta^{*}$ is even more pronounced in
the geometrically averaged local density of states (see Appendix \ref{sec:tdos}).

\subsection{Random coupling to superconducting substrate}

\begin{figure} 
\includegraphics[width=0.85\linewidth]{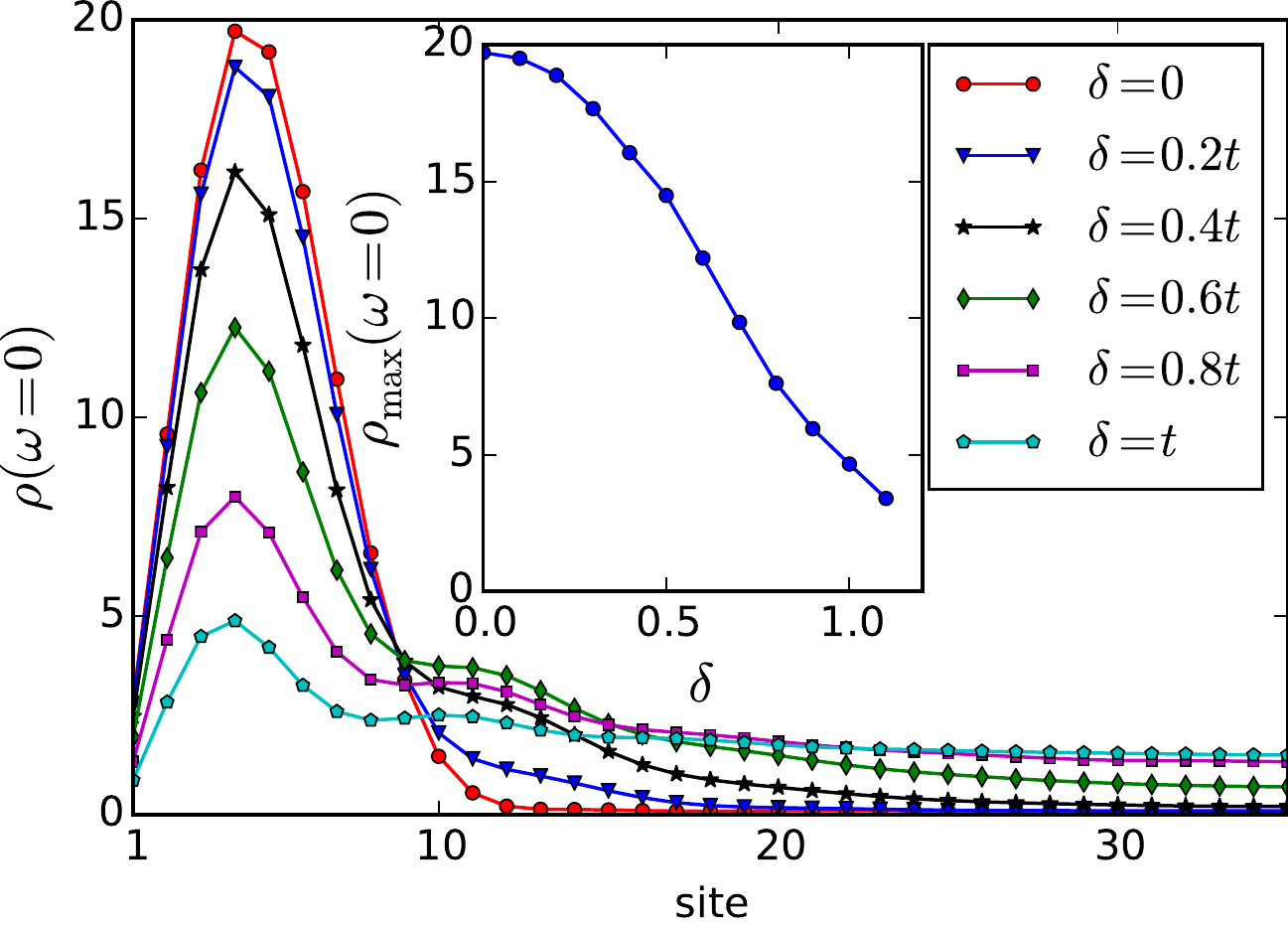}
\caption{Evolution of the subgap quasiparticle spectrum 
$\rho_{i}(\omega=0)$ caused  by the random coupling $\Gamma_{i,S}$.} 
\label{offdiag_disorder}
\end{figure}
%

In analogy to the random site energies (discussed in previous subsection) 
we have studied the inhomogeneous coupling between the chain atoms and 
superconducting substrate $\Gamma_{i,S}=\Gamma_{S} + \xi_{i}\,\delta$. 
Physically such situation may occur when the wave-functions of atoms 
randomly overlap with the wave-functions of itinerant electrons 
in superconducting substrate. This kind of disorder (that appears in
off-diagonal Green's function of the Nambu representation) is known 
to have detrimental influence on superconductivity of bulk materials 
\cite{BalatskyRMP06}. 

Fig.\ \ref{offdiag_disorder} shows the results obtained for several amplitudes $\delta$ 
(in units of $t$) of the off-diagonal disorder imposed on the reference 
value $\Gamma_{S}/t=0.2$. We roughly observe a similar tendency as in 
Fig.\ \ref{diag_disorder}, but a more careful examination indicates that 
nonuniform coupling $\Gamma_{i,S}$ is slightly less influential on the 
Majorana quasiparticles \cite{Wang-2016}.

\subsection{Topological quantum number\label{topo}}
 
 So far we have analyzed evolution of the local density of states 
 at $\omega = 0$, providing indirect evidence for the disorder--induced 
 destruction of the Majorana bound states. We argued that transfer 
 of the averaged spectral density $\rho(\omega)$ from the chain ends 
 to its internal lattice sites  (below some critical $\delta^{*}$) 
 was caused by pinning of the Majorana states by disorder. For  stronger 
 $\delta > \delta^{*}$ the Majorana states disappeared and 
 the quasiparticle peaks moved to finite (non-zero) energies. In consequence 
 the averaged $\rho(\omega=0)$ decreased and its maximal value (located at 
 the central chain sites) indicated a smooth transition to the topologically 
 trivial superconducting phase. In order to check whether this interpretation 
 of transition from the topologically nontrivial to trivial regions 
 (at $\delta^*$) is correct we shall analyze it here by  determining 
 the $\mathbb{Z}_2$ topological quantum number.

 The topological quantum number ${\cal Q}$ can identify whether the chain 
 is in the topologically trivial (${\cal Q}=1$) or nontrivial (${\cal Q}=-1$) 
 state. Kitaev\cite{Kitaev-2001} has shown that for a translationally 
 invariant nanowire this quantity can be expressed as ${\cal Q}={\rm 
 sign\,Pf}[A(0)]{\rm Pf}[A(\pi)]$, where $A(k)$ is the Hamiltonian in 
 momentum space transformed to the Majorana basis. In the disordered systems, 
 however, skew-symmetric matrices $A(k)$ are very large so it is 
 cumbersome to calculate ${\cal Q}$ from the above formula.
  
\begin{figure}[H] 
\includegraphics[width=\linewidth]{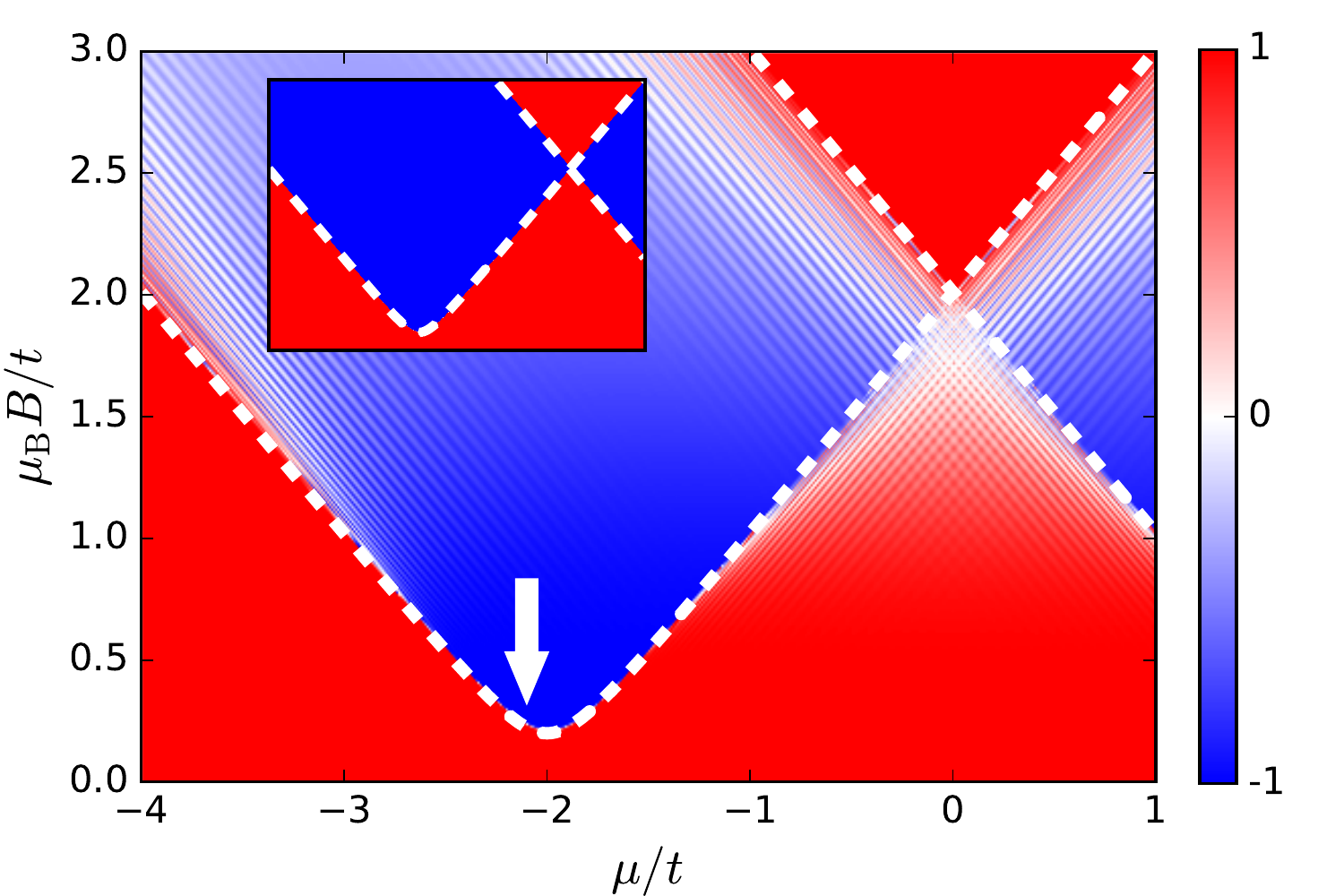}
\caption{$\det (R)$ as a function of the chemical 
potential $\mu$ and magnetic field $B$ for a clean chain composed of 70 
lattice sites. The rest of the model parameters are the same as assumed in 
the calculations.
The white dashed lines show the boundaries of the topologically nontrivial 
regime given by Eq. (\ref{eq:boundaries}). The arrow indicates 
the parameters used in the study of the disorder--induced destruction 
of the Majorana end states.
The inset shows the same for a 500-site chain.}
\label{fig:Q1}
\end{figure}

\begin{figure}[H] 
\includegraphics[width=\linewidth]{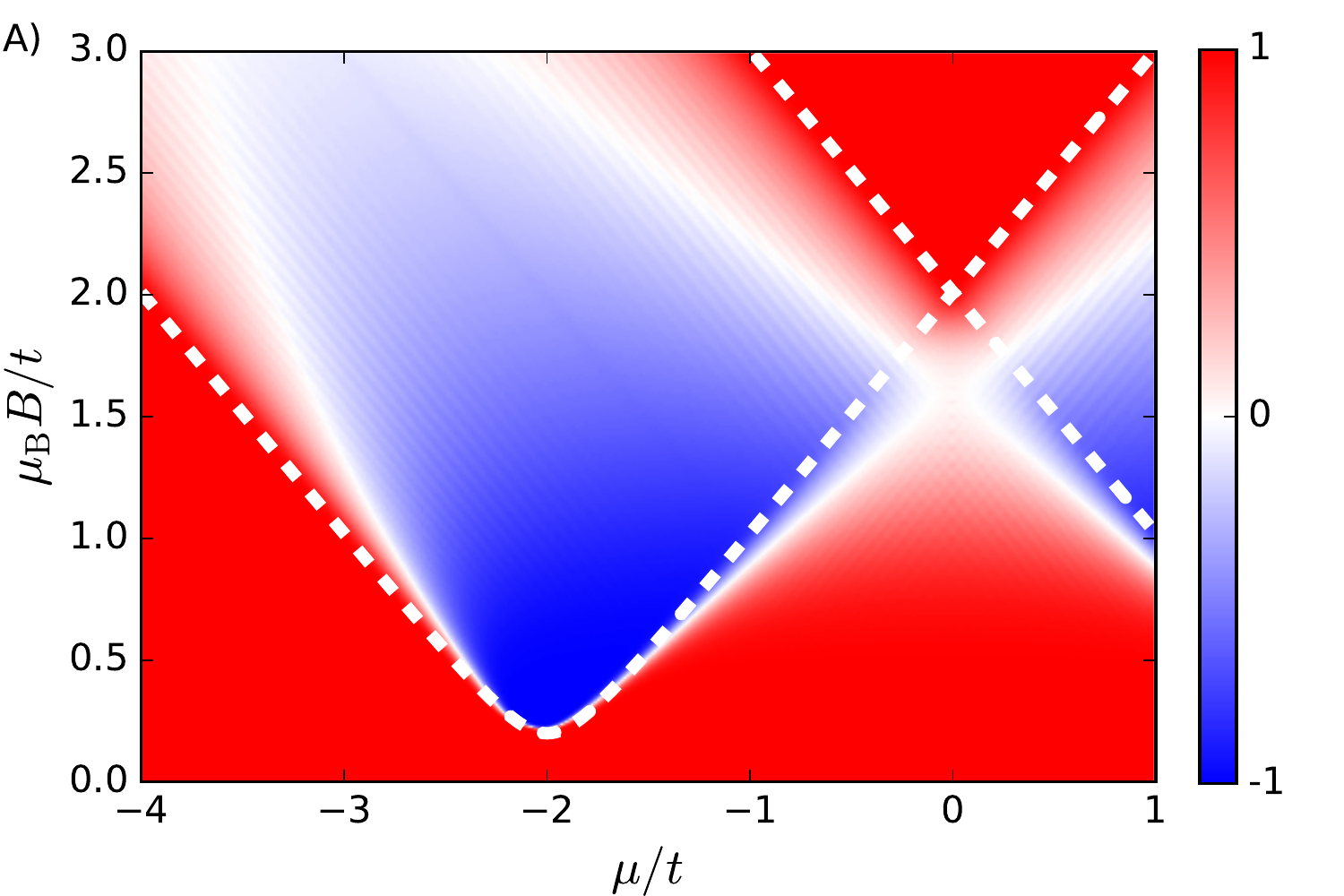}
\includegraphics[width=\linewidth]{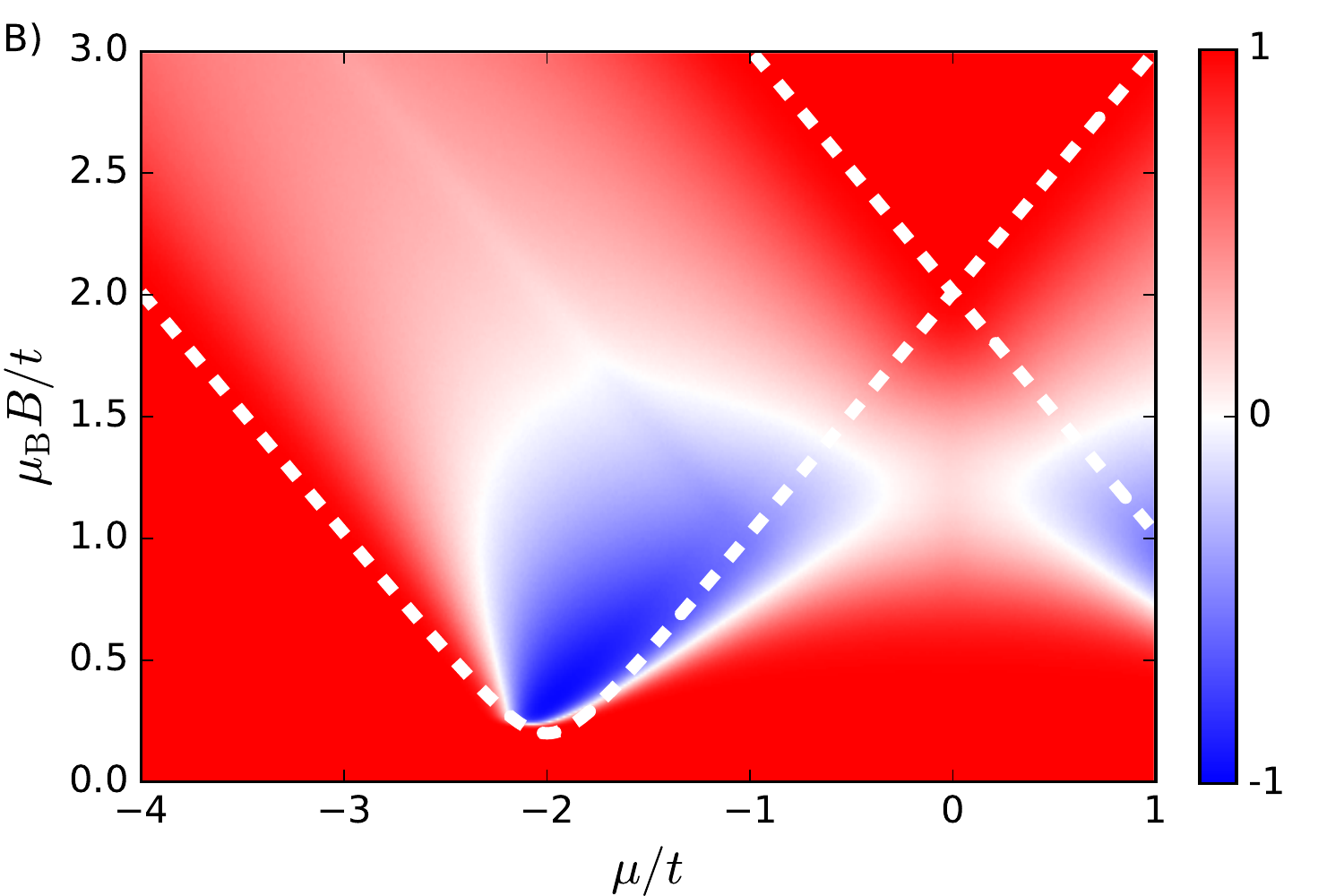}
\caption{The same as in Fig. \ref{fig:Q1}, but in the presence of disorder 
with $\delta=0.25t$ (A) and $\delta=0.5t$ (B), respectively. Value of $\det (R)$ 
is averaged over $10^4$ realizations of disorder. The white area separating 
the red and blue regions correspond to border-line (where $\det (R)$ changes 
sign) between the topologically distinct phases.}
\label{fig:Q2}
\end{figure}

In Refs.\ \onlinecite{Akhmerov-2011,Fulga-2011} it has been shown that 
the topological number ${\cal Q}$ can be determined from the scattering 
matrix $S$ of the chain
  \begin{equation}
    S=\left(\begin{array}{cc}
      R & T'\\
      T & R'
    \end{array}\right),
    \label{matrix_s}
  \end{equation}
  where $R$ and $T$ ($R'$ and $T'$) are $4\times 4$ reflection and transmission matrices, respectively, at the left (right) end of the chain. This matrix describes transport through the chain coupled to left and right lead
  \begin{equation}
    \left(\begin{array}{c}\psi_{-,{\rm L}} \\ \psi_{+,{\rm R}}\end{array}\right)
    =S\left(\begin{array}{c}\psi_{+,{\rm L}} \\ \psi_{-,{\rm R}}\end{array}\right),
  \end{equation}
  where $\psi_{\pm,L/R}$ are the right or left moving modes ($\pm$) at the left or right edge ($L/R$) at the Fermi level. Then, the topological quantum number is given by
  \begin{equation}
      {\cal Q}={\rm sign\,det}(R) = {\rm sign\,det}(R').
  \end{equation}
  The scattering matrix $S$ can be obtained from multiplication of the individual transfer matrices of all the lattice sites. This procedure 
  has been described in detailed in Ref. \onlinecite{Choy-2011}. Since 
  the product of numerous transfer matrices is numerically unstable, we converted them into a composition of the unitary  matrices,  involving 
  only eigenvalues of unit absolute value. This stabilization method has
been proposed in Ref. \onlinecite{Snyman-2008}.
   Fig. \ref{fig:Q1} shows $\det (R)$ as a function 
  of the chemical potential $\mu$ and magnetic field $B$ 
  for the uniform Rashba chain (without any disorder).
%
Difference between the boundaries of the topologically nontrivial regime
given by Eq. (\ref{eq:boundaries}) and the actual change of sign of $\det (R)$
results from the finite length of Rashba chain. By comparing the main plot with 
the inset in Fig. \ref{fig:Q1} (corresponding to 500-site chain) we clearly
observe, that topological properties for the model parameters (chosen in Sec. II) 
are only slightly smeared by the finite--size effects. In such regime 
the topologically nontrivial state remains intact also 
in presence of the  weak disorder ($\delta < \delta^*$). This can be
seen in Fig. \ref{fig:Q2}, where we display  $\det(R)$ 
averaged over $10^4$ disorder realizations for  $\delta=0.25t$ and 
$\delta=0.5t$, respectively.

Criterion based on the topological number ${\cal Q}$ was used by 
us to establish the magnitude of disorder, above which the Majorana 
bound states disappear. Fig. \ref{middle_point}B shows that such 
topological transition indeed takes place around $\delta^{*}$. 
This confirms that maximum of $\rho(\omega=0)$ in the central part 
of the Rashba chain coincides with the crossover between the topologically
distinct phases. Above $\delta^*$ the system enters the topologically 
trivial phase, where the bound quasiparticle states are lifted to 
finite (non-zero) energies, leading to suppression of $\rho(\omega=0)$. 
 A closer look at Fig. \ref{middle_point}B reveals that $\det(R)$ in the
 changeover regime is getting steeper with increasing length of the chain
 and the point where it changes its sign is moving towards stronger 
 disorder. The critical disorder $\delta^*$ in the limit of an infinite
 chain is indicated by the point where all the lines representing $\det(R)$
 for different chain lengths cross each other. This point is very close 
 to the maximum of $\rho(\omega=0)$.

\section{Single impurities\label{sec:partitioning}}

Another interesting situation can be caused by single impurities
existing either in the Rashba chain or attached to it. We briefly
address such problem in this section. 

%
\begin{figure}[!htb] 
\includegraphics[width=0.8\linewidth]{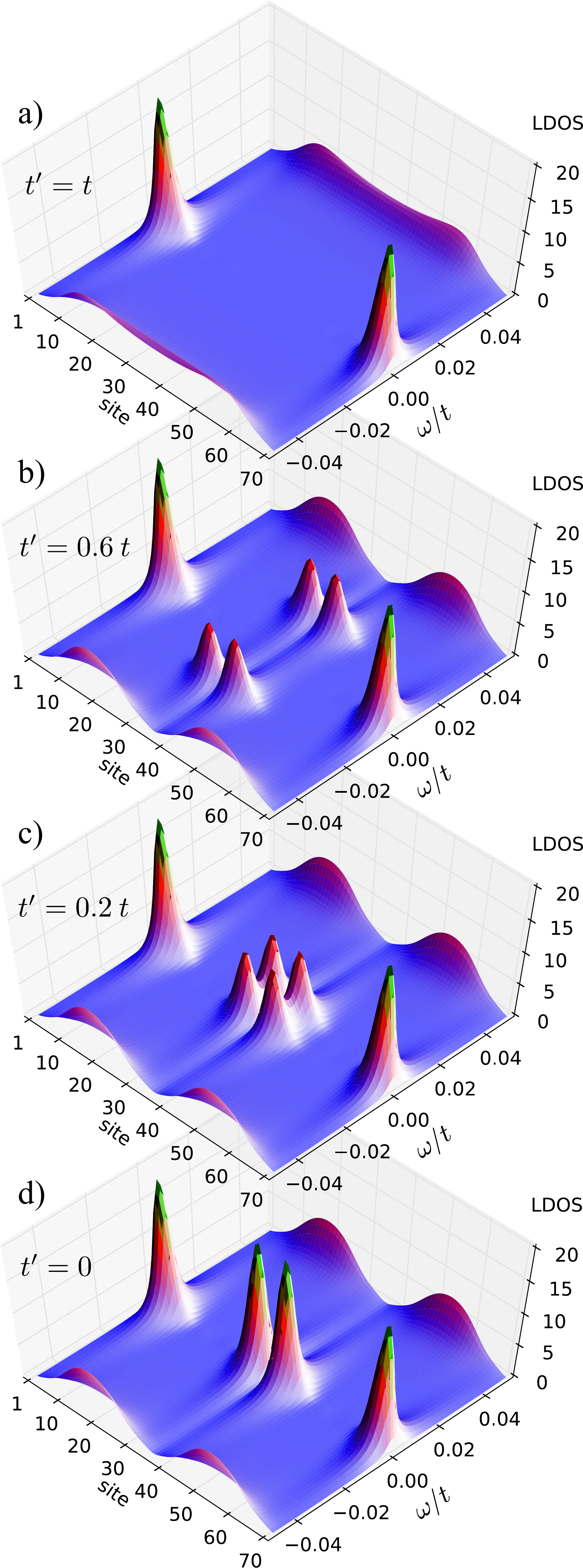}
\caption{The spatially and energy dependent spectral function 
$\rho_{i}(\omega)$ of the atomic chain, where the hopping
integral between the internal 35-th and 36-th sites is reduced 
to $t'/t=$1, 0.6, 0.2, 0 (from top to bottom).}
\label{fig7}
\end{figure}
%

\subsection{Splitting of the Rashba chain}

Let us consider a single internal `defect', that can effectively produce 
additional pair of the Majorana quasiparticles. To be specific, we 
assume a reduced hopping integral $t'$ between one pair of the internal 
sites in the Rashba chain. For numerical calculations we imposed 
the reduced hopping $t'$ between 35$^{\mbox{\footnotesize \rm th}}$ 
and 36$^{\mbox{\footnotesize \rm th}}$ sites of the atomic chain, 
consisting of 70 atoms. We checked, that a particular position 
of the reduced interatomic hopping is not crucial as long as it is 
well in-between the Majorana quasiparticles.

\begin{figure}[H] 
\includegraphics[width=0.385\linewidth, angle=90]{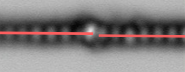}
\caption{Constant-height zero-bias AFM image of the Fe chain 
used in the Basel experiment. The red lines drawn on top of the
image show that there are sections shifted in the direction perpendicular 
to the chain. This defect image is part of Fig. 2c  
from Ref.\ \onlinecite{Kisiel-15}
[R. Pawlak {\em et al.} ``Probing Atomic Structure and Majorana Wavefunctions in Mono-Atomic Fe-chains on Superconducting Pb-Surface'', npj Quantum Information {\bf 2} 16035, (2016)] used 
and modified in accordance with the Creative Commons Attribution (CC BY) license.
\label{wire_kisiel}}
\end{figure}

In Fig.\ \ref{fig7} we illustrate emergence of the Majorana 
quasiparticles driven by the suppressed hopping integral $t'<t$ 
(which can be regarded as a particular kind of the local defect). 
In the limit $t' \rightarrow 0$ the Rashba chain is decomposed into 
separate segments and during this process one pair of the finite-energy
(Andreev) states gradually evolves into the zero-energy Majorana states. 
Although this phenomenon is rather obvious, it is quite astonishing 
that the Majorana states, already existing at the edges of uniform chain, 
are completely unaffected by the chain partitioning. We assign 
this unique behavior to non-local character of the 
edge Majorana states. This effect is similar 
to what we presented in Sec.\ \ref{fse}, investigating 
the Majorana quasiparticles of the atomic chain in absence 
of the spin--orbit coupling on its interior part.
Similar internal pair of the Majorana quasiparticle states have 
been  predicted  on interface between the chain segments, 
where the spin--orbit coupling $\alpha$ changes sign \cite{Klinovaja-2015}.
The  reduced hopping $t'$ considered here seems to be quite realistic 
because it could explain origin of the bright spots observed near 
the locally deformed Fe-atom chain in the Basel group experiment 
(Fig. \ref{wire_kisiel}). To verify this, one should inspect 
the spatially resolved differential conductance measured 
around such defect. 
%

\subsection{Diagonal and off-diagonal impurity}

%

Similar effects can be also achieved if we place either the diagonal 
or off-diagonal impurity in one of the internal Rashba chain sites. 
In practice, however, such `internal impurity' has to be extremely distinct from all
remaining constituents of the chain. 
Our numerical calculations showed that the impurity potential 
has to be very large ($\delta > 10 t$) to effectively break 
the chain into separate pieces, generating an additional 
pair of the Majorana quasiparticles.
This situation is hardly probable in realistic systems, 
therefore we discard it from present considerations.

\subsection{Side-attached normal impurity}

Let us finally analyze  what happens when a (normal) quantum impurity
is coupled to the Rashba chain at the sites, which host the Majorana 
quasiparticle. Similar issue has been previously addressed, 
considering the impurity coupled to the very end of Kitaev chain 
\cite{Klinovaja-2012,Chevallier-2012,Vernek-2014,Vernek-2015}.
In finite size Rashba chain the maximum probability of the Majorana 
quasiparticles does not coincide with the very last sites, but they are 
located around 4$^{\mbox{\footnotesize \rm th}}$ site from the atomic edge 
(as  reported experimentally \cite{Yazdani-2015}). For this reason 
we expect the strongest influence of the Majorana quasiparticle
on the normal quantum impurity when the latter is side-attached
to 4$^{\mbox{\footnotesize \rm th}}$ site of the Rashba chain.

\begin{figure} 
\includegraphics[width=\linewidth]{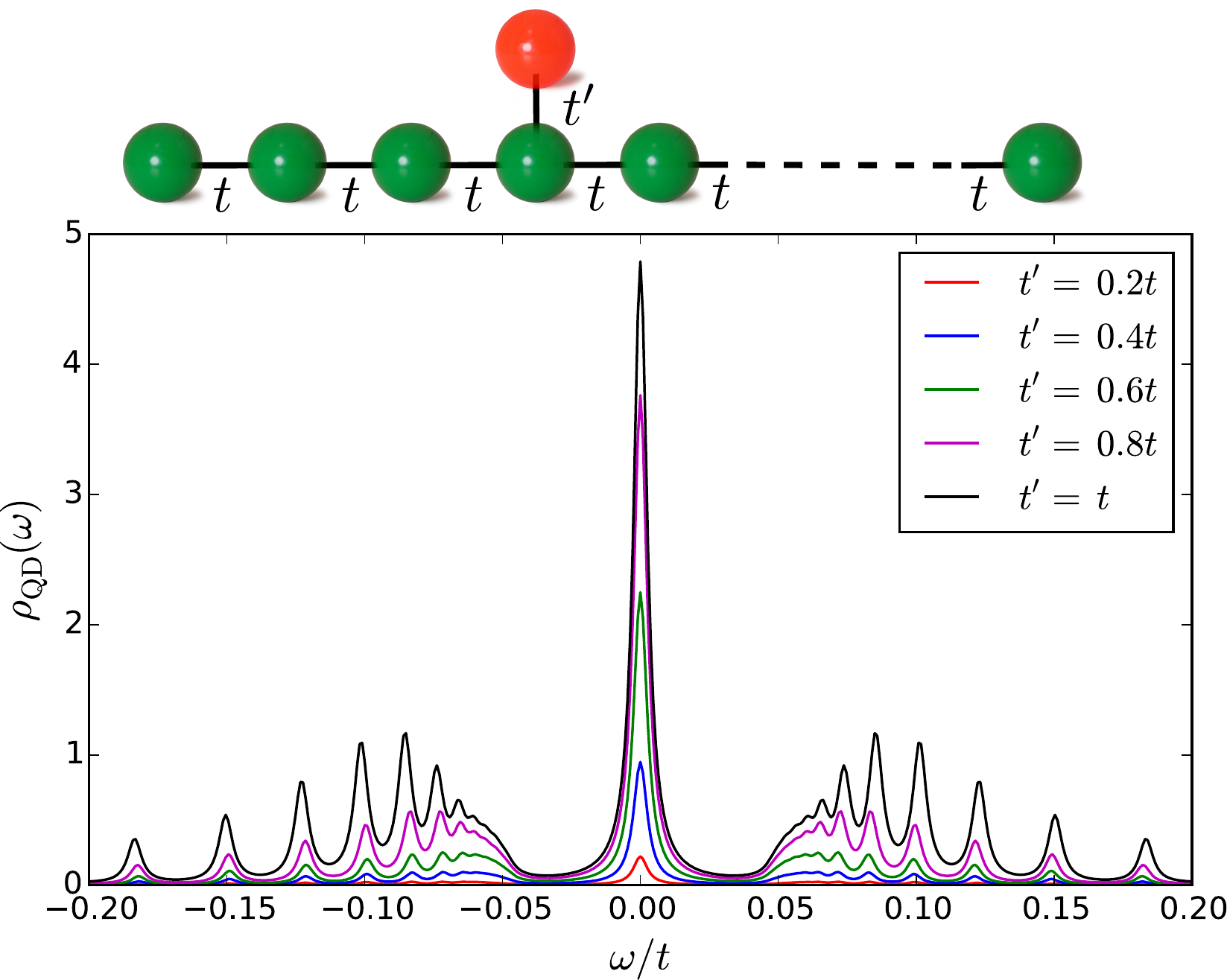}
\caption{The proximity effect showing up in the spectrum of 
the normal quantum impurity (the red sphere in the top panel) that 
is side-attached to the atomic chain, near the Majorana quasiparticle.}
\label{proximity_effect}
\end{figure}

On a formal level, we studied the model Hamiltonian (\ref{proximized})
supplemented with the term
\begin{eqnarray}
H_{imp}=\varepsilon_{f} \sum_{\sigma} \hat{f}_{\sigma}^{\dagger} 
\hat{f}_{\sigma} + t_{f} \sum_{\sigma} \left( \hat{f}_{\sigma}^{\dagger} 
\hat{d}_{j,\sigma} +\mbox{\rm h.c.} \right) 
\label{normal_impurity}
\end{eqnarray}
describing the quantum impurity of energy $\varepsilon_{f}$ attached 
to site $j=4$ of the topological wire. In Fig.\ \ref{proximity_effect}
we present electronic spectrum of the quantum impurity $\rho_{f}(\omega)$
obtained for several couplings $t_{f}/t$, as indicated. 
With increasing  $t_{f}$ we observe appearance of the zero-energy state, 
that absorbs more and more spectral weight. This feature is caused by 
the proximity effect, which can be considered  as `leakage' of Majorana 
mode onto normal quantum impurities  \cite{Vernek-2014,Vernek-2015}.
As a minor remark, let us notice that the finite energy (Andreev/Shiba) states do leak into the normal quantum impurity as well.

Summarizing this section, we conclude that various types of quantum 
impurities may play important role for the Majorana quasiparticles of
the Rashba chain. The reduced hopping integral or the strong local
potential (both in the single particle and pairing channels) can
effectively break the chain into pieces, so that additional pair 
of Majorana quasiparticles emerges. On the other hand, any nanoscopic
object coupled to the existing Majorana quasiparticle absorbs such 
exotic entity in exactly the same way as the superconducting or 
magnetic order spreading onto some neighboring normal regions.  

\section{Summary}

We studied the subgap electronic spectrum of the inhomogeneous Rashba 
chain deposited on $s$-wave superconducting substrate, where the strong 
spin--orbit coupling combined with the Zeeman effect induce the zero-energy 
Majorana quasiparticles near its edges. We analyzed the spatial extent
of such quasiparticles due to intrinsic inhomogeneity caused by the 
boundary effects. In accordance with the STM experimental observations 
\cite{Yazdani-2015} the zero-bias enhancement is mostly pronounced 
by the Majorana quasiparticle centered at either the 
4$^{\mbox{\footnotesize \rm th}}$ or 3$^{\mbox{\footnotesize \rm rd}}$ 
atom from the chain edge. 

We also investigated stability of the Majorana quasiparticles against
various types of the disorder, that are likely to occur in realistic
STM-type configurations (schematically displayed in Fig.\ \ref{fig1}). 
Despite a common belief, that the Majorana quasiparticles are robust 
to environmental influence we have shown that this is not truly the 
case. Our study reveals that sufficiently strong disorder would be 
detrimental for the Majorana quasiparticles, causing a transition 
from the topologically nontrivial to trivial superconducting phases.
This conclusion is unambiguously supported by the 
value of the $\mathbb{Z}_2$ topological number, averaged over numerous
($\sim 10^{4}$) different configurations with the fixed amplitude $\delta$.
We have shown that such a qualitative changeover would be 
spectroscopically manifested by a peak in the zero-energy spectrum 
in the internal chain sites (Fig.\ \ref{middle_point}). 
We demonstrated that this peak is present not only in a simple 
single--band model, but also in a more realistic multiband one.

Since it is difficult to measure the topological invariant, this effect
might be useful for empirical determination of the critical 
disorder $\delta^{*}$, signalling a transition between
the non--trivial and trivial phases. It would particularly 
well suited for systems, where disorder can be modified in a controllable
way, e.g.\ in ultracold atoms\cite{controllable_disorder}.

Single quantum impurities have also very unusual interplay with the Majorana 
quasiparticle states. Under specific conditions they can effectively induce 
additional pairs of the Majorana states, when the strong impurity scattering
affects any internal site of the Rashba chain. This process is analogous to 
partitioning the chain into separate pieces, bringing to life new Majoranas.
It is amazing, however, that the existing (external) Majorana modes are
practically left intact by such partitioning. This unusual phenomenon is 
a signature of their non-local origin that could be useful for constructing 
the quantum bits out of Majorana quasiparticles. We also have shown that
Majorana quasiparticles can spread onto nanoscopic objects 
coupled to them. Such proximity effect can be helpful for designing some
novel tunneling hetero-structures to indirectly probe the Majorana 
quasiparticles, for instance by the quantum interference.

\section*{Acknowledgments:}

We thank M.~Kisiel for discussions and pointing to us  signatures of the internal chain defect reported 
in Ref.~\onlinecite{Kisiel-15}.
This work is supported by the National Science Centre under the contracts 
DEC-2014/13/B/ST3/04451 (TD) and DEC-2013/11/B/ST3/00824 (MMM).

\appendix
\section{In-gap states of the proximized atom \label{sec:uncorQD}}
\label{A}

To support relevance of the microscopic model (\ref{proximized}) for the deep 
subgap regime ($|\omega|\ll\Delta$) let us study influence of the proximity 
effect on the single atom (Anderson-type impurity) placed between the normal
and superconducting reservoirs, neglecting the inter-site 
hopping $t_{ij}=0$. Since all atoms behave identically we can skip index $i$
when studying the spectrum of a given atom. To account for the proximity effect 
we have to treat the particle and hole degrees of freedom on equal footing. 
This can be done within the Nambu representation $\hat{\Psi}_{d}^{\dagger}
=(\hat{d}_{\uparrow}^{\dagger},\hat{d}_{\downarrow})$, $\hat{\Psi}_{d}=
(\hat{\Psi}_{d}^{\dagger})^{\dagger}$ introducing the single particle 
matrix Green's function ${\mb G}(\tau,\tau') \!=\! \langle\!\langle 
\hat{\Psi}_{d}(\tau); \hat{\Psi}_{d}^{\dagger}(\tau')\rangle\!\rangle$. 
The electronic spectrum and the transport properties (see Appendix B) 
are given by its diagonal and off-diagonal parts, respectively.

In equilibrium conditions ($\mu_{N}=\mu_{S}$) the Fourier transform of 
${\mb G}_{d}(\tau,\tau')={\mb G}_{d}(\tau-\tau')$ can be expressed as
\begin{eqnarray} 
{\mb G}^{-1}(\omega) = 
\left( \begin{array}{cc}  
\omega\!-\!\varepsilon &  0 \\ 0 &  
\omega\!+\!\varepsilon\end{array}\right)
- {\mb \Sigma}_{d}(\omega) .
\label{GF}\end{eqnarray} 
In general, the self-energy ${\mb \Sigma}_{d}(\omega)$ accounts for the hybridization 
of the quantum impurity with external reservoirs and the correlation effects. 
In absence of the correlations its exact form  
${\mb \Sigma}_{d}(\omega)=\sum_{{\bf k},{\beta}}|V_{{\bf k}\beta}|^{2}
g_{{\bf k}\beta}(\omega)$ depends on the Green's functions 
$g_{{\bf k}\beta}(\omega)$ of mobile electrons. 
In the wide-band limit ${\mb \Sigma}_{d}(\omega)$ simplifies 
to \cite{Bauer-07,Yamada-11} 
\begin{eqnarray}
\mb{\Sigma}_{d}(\omega) &=&  
- \; \frac{i \Gamma_{N}}{2} \; \left( \begin{array}{cc}  
1 & 0 \\ 0 & 1 \end{array} \right)
 -  \frac{\Gamma_{S}}{2} \left( \begin{array}{cc}  
1 & \frac{\Delta}{\omega} \\ 
 \frac{\Delta}{\omega}  & 1 
\end{array} \right) \nonumber \\
& \times &
\left\{
\begin{array}{ll} 
\displaystyle\frac{\omega}{\sqrt{\Delta^{2}-\omega^{2}}}
& \mbox{\rm for }  |\omega| < \Delta  \\
\displaystyle\frac{i\;|\omega|}{\sqrt{\omega^{2}-\Delta^{2}}}
& \mbox{\rm for }  |\omega| > \Delta 
\end{array} \right. . 
\label{selfenergy_0}
\end{eqnarray} 
Eq. (\ref{selfenergy_0}) describes: (i) the proximity 
induced electron pairing (via the off-diagonal terms that are proportional 
to $\Gamma_{S}$) and (ii) the finite life-time effects. The latter 
come from the imaginary parts of the self-energy (\ref{selfenergy_0})
and depend either on both couplings $\Gamma_{\beta =N,S}$ 
(for energies $|\omega|\geq\Delta$) or solely on $\Gamma_{N}$ 
(in the subgap regime  $|\omega|<\Delta$).

%
\begin{figure} %
\includegraphics[width=0.85\linewidth]{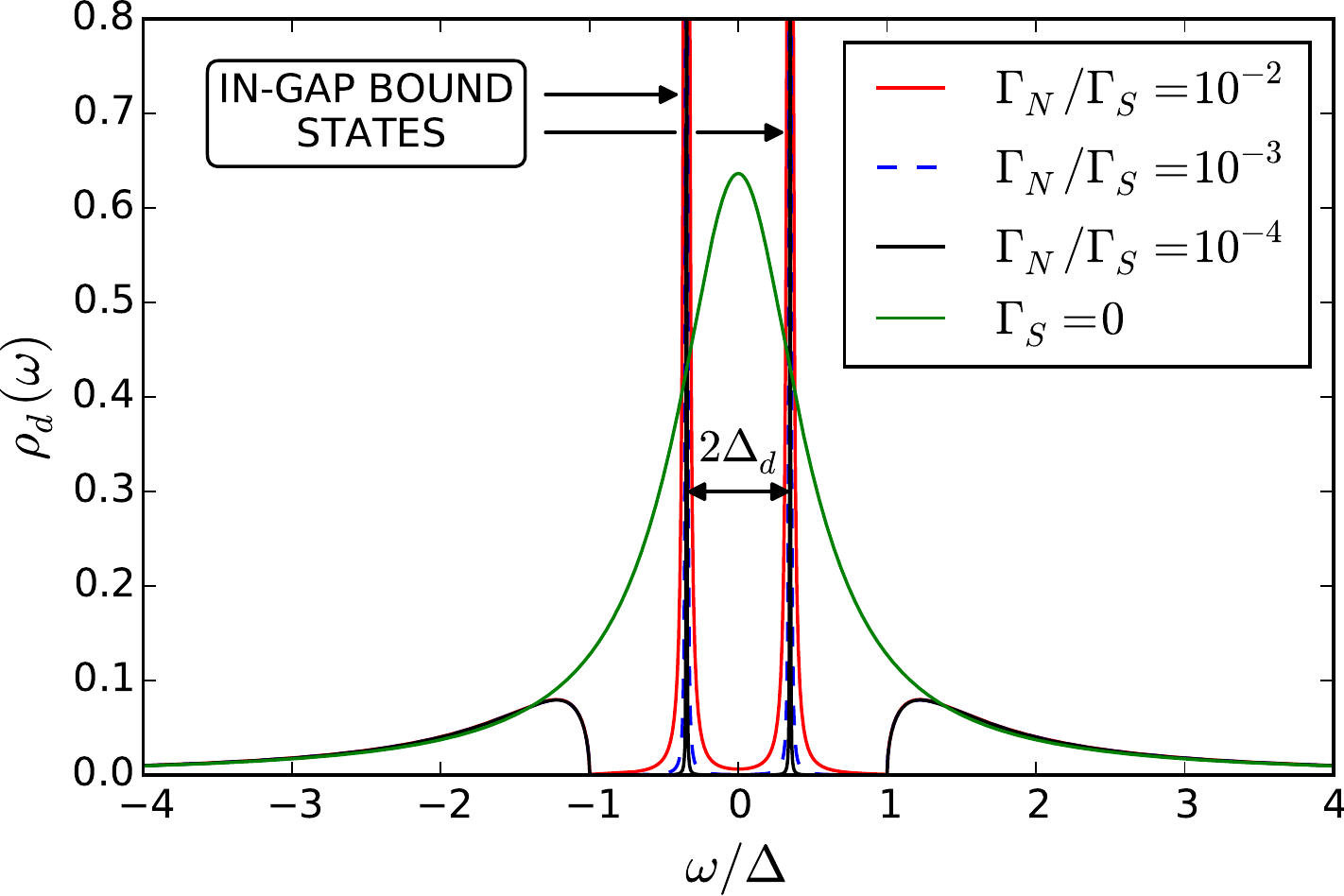}
\caption{Spectrum of the isolated atom ($t_{ij}=0$) coupled to 
the superconducting ($\Gamma_{S}$) and normal ($\Gamma_{N}$) reservoirs.
Results are obtained in absence of the correlations for $\varepsilon=0$, 
$\Delta=2\Gamma_{S}$ and several coupling ratios $\Gamma_{N}/\Gamma_{S}$, 
as indicated.}
\label{fig10}
\end{figure}
%

In the subgap regime $|\omega|<\Delta$ the Green's function 
acquires the BCS-type structure
\begin{eqnarray}
{\mb G}(\omega) &=& 
\left( \begin{array}{cc}  
\tilde{\omega} + i \Gamma_{N}/2 - \varepsilon   
\hspace{0.3cm}&  \tilde{\Gamma}_{S}/2 \\ 
\tilde{\Gamma}_{S}/2  & 
\tilde{\omega} + i \Gamma_{N}/2 + \varepsilon   
\end{array} \right)^{-1} 
\label{G_0}
\end{eqnarray}
with  $\tilde{\omega} = \omega + \frac{\Gamma_{S}}{2} \frac{\omega}
{\sqrt{\Delta^{2}-\omega^{2}}}$ and $\tilde{\Gamma}_{S} = \Gamma_{S} 
\frac{\Delta}{\sqrt{\Delta^{2}-\omega^{2}}}$, respectively. The spectral 
function $\rho_{d}(\omega)=-\pi^{-1}\mbox{\rm Im}{\mb G}_{11}(\omega+i0^{+})$
reveals two in-gap peaks, related to the Andreev \cite{Bauer-07,Andreev} 
or Yu-Shiba-Rusinov \cite{Yu-Shiba-Rusinov,BalatskyRMP06} quasiparticles. 
We can regard their splitting as the induced pairing gap $\Delta_{d}$  
(Fig.\ \ref{fig10}).

Fig.\ \ref{fig10} shows the spectral function for several ratios $\Gamma_{N}/\Gamma_{S}$.
In the extreme regime $\Gamma_{N}\!\rightarrow\!0$ the in-gap quasiparticles
are represented by the Dirac distribution functions (i.e.\ their lifetime 
is infinite). Otherwise the broadening of in-gap states is proportional to 
$\Gamma_{N}$. Energies $E_{A,\pm}$ of the Andreev quasiparticles have to be 
determined from the following equation \cite{Baranski-13}
\begin{eqnarray}
E_{A,\pm} +  \frac{(\Gamma_{S}/2)E_{A,\pm}} {\sqrt{\Delta^{2}
-E_{A,\pm}^{2}}} = \pm \sqrt{\varepsilon^{2}+ 
\frac{(\Gamma_{S}/2)^{2}\Delta^{2}}{\Delta^{2}-E_{A,\pm}^{2}}} .
\label{energy_eqn}
\end{eqnarray}
For $\Gamma_{S}\gg \Delta$ the quasiparticle energies 
(\ref{energy_eqn}) appear close to the superconductor gap edges $E_{A,\pm}\simeq 
\pm\Delta$, whereas in the weak coupling limit, $\Gamma_{S}\ll 
\Delta$, the asymptotic values are $E_{A,\pm} \simeq \pm 
\sqrt{\varepsilon+\left(\Gamma_{S}/2\right)^{2}}$. 

In the 'superconducting atomic' limit $\Gamma_{N}\rightarrow 0$ 
the self\-energy (\ref{selfenergy_0}) becomes static
\begin{eqnarray}
\mb{\Sigma}_{d}^{0}(\omega) &=&  - \; \frac{1}{2} \; 
\left( \begin{array}{cc}  i\Gamma_{N} & \Gamma_{S} \\ 
\Gamma_{S} & i\Gamma_{N} \end{array} \right)
\end{eqnarray}
therefore the following equivalence holds
\begin{eqnarray}
&&\sum_{\sigma}  \varepsilon \hat{d}^{\dagger}_{\sigma} 
\hat{d}_{\sigma}+ \hat{H}_{S} + \sum_{{\bf k},\sigma}  
\left( V_{{\bf k} S} \; \hat{d}_{i,\sigma}^{\dagger}  
\hat{c}_{{\bf k} \sigma S} + \mbox{\rm h.c.} \right)
\nonumber \\
& = & \sum_{\sigma}  \varepsilon \hat{d}^{\dagger}_{\sigma} 
\hat{d}_{\sigma} - \left( \Delta_{d} \hat{d}_{\uparrow}^{\dagger}
\hat{d}_{\downarrow}^{\dagger} + \mbox{\rm h.c.} \right)
\label{equivalence} 
\end{eqnarray}
with $\Delta_{d}=\Gamma_{S}/2$  valid for the deep subgap 
regime $|\omega| \ll \Delta$. This line of reasoning allows 
us to represent the initial Hamiltonian (\ref{model}) of 
the atomic chain by (\ref{proximized}) that describes 
the 'proximized' quantum wire coupled to the STM tip.

\section{In-gap states of the chain \label{sec:ingapstates}}
\label{B}

\begin{figure} 
\includegraphics[width=0.8\linewidth]{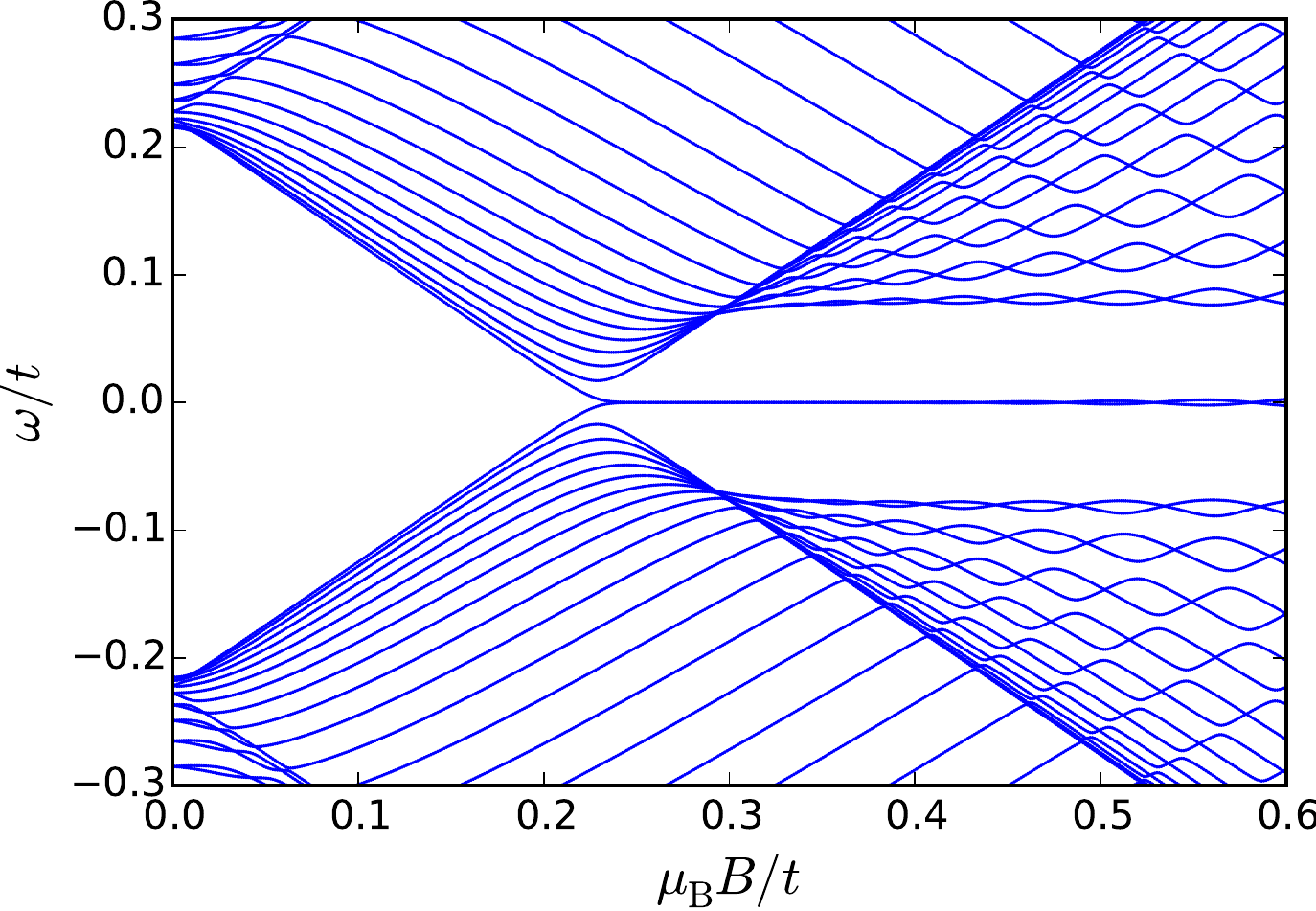}
\caption{Evolution of the subgap quasiparticle energies of the uniform 
Rashba chain from the (topologically) trivial to nontrivial 
superconducting phases above $B_{c}\approx 0.22$.}
\label{fig2}
\end{figure}
%

In Fig.\ \ref{fig2} we show typical variation of the in-gap quasiparticle energies 
versus the magnetic field $B$, expressed in units of $t/(g\mu_{B}/2)$.
We notice that at some critical value ($B_{c} \approx 0.22$) two in-gap
quasiparticles evolve into the zero-energy bound states. Because of 
a finite chain-length there is some overlap between these Majoranas, 
observable by a tiny splitting of the zero-energy modes with magnitude 
dependent on the magnetic field. Above $B_{c}$ these Majorana modes are 
protected by the soft gap ($\sim 0.1 t$ for the present set of parameters) 
outside of which there exist the ordinary (Andreev/Shiba) states.

\section{Crossover to trivial phase \label{sec:tdos}}

\begin{figure}[H]
\includegraphics[width=0.8\linewidth]{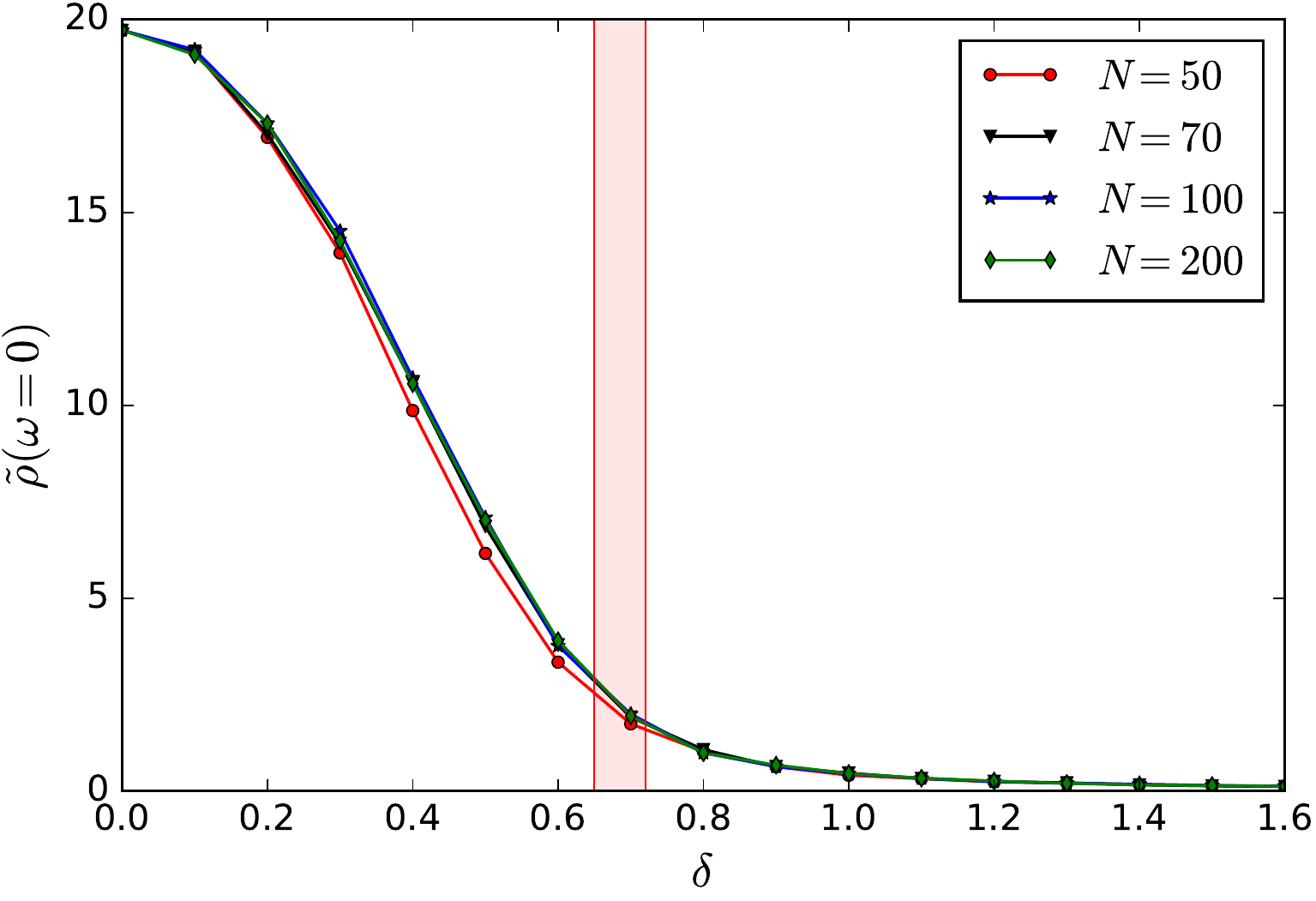}
\caption{Geometrically averaged  density of states obtained at 
$\omega=0$ for site $i=4$ as a function of the diagonal disorder. 
The red vertical strip corresponds to the
changeover from the nontrivial to trivial superconducting 
phases (as defined in Fig.~\ref{middle_point})
\label{fig:tdos_koniec}}
\end{figure}

In this section we provide additional evidence for a  crossover 
from the nontrivial ($p$-wave) to trivial (s-wave) pairing driven by the
diagonal disorder (random energies of the atoms). Fig.\ 
\ref{fig:tdos_koniec} shows the geometrically averaged density of states 
$\tilde{\rho}_{i}(\omega)\equiv \exp \langle \ln{\rho_{i}(\omega)} 
    \rangle_{\{ \xi_{i} \} }$ ({\em typical density of states}) versus the disorder amplitude $\delta$ 
at site $i=4$, corresponding  to the maximum of the zero-energy  
Majorana mode. We noticed that length $N$ has practically no effect, 
therefore we conclude that gradual disappearance of the Majorana 
quasiparticles is a generic property. Point $\delta^{*}$ 
can be interpreted as characteristic for the smooth transition from 
the nontrivial to trivial superconducting phases.

\begin{figure}[H]
\includegraphics[width=0.8\linewidth]{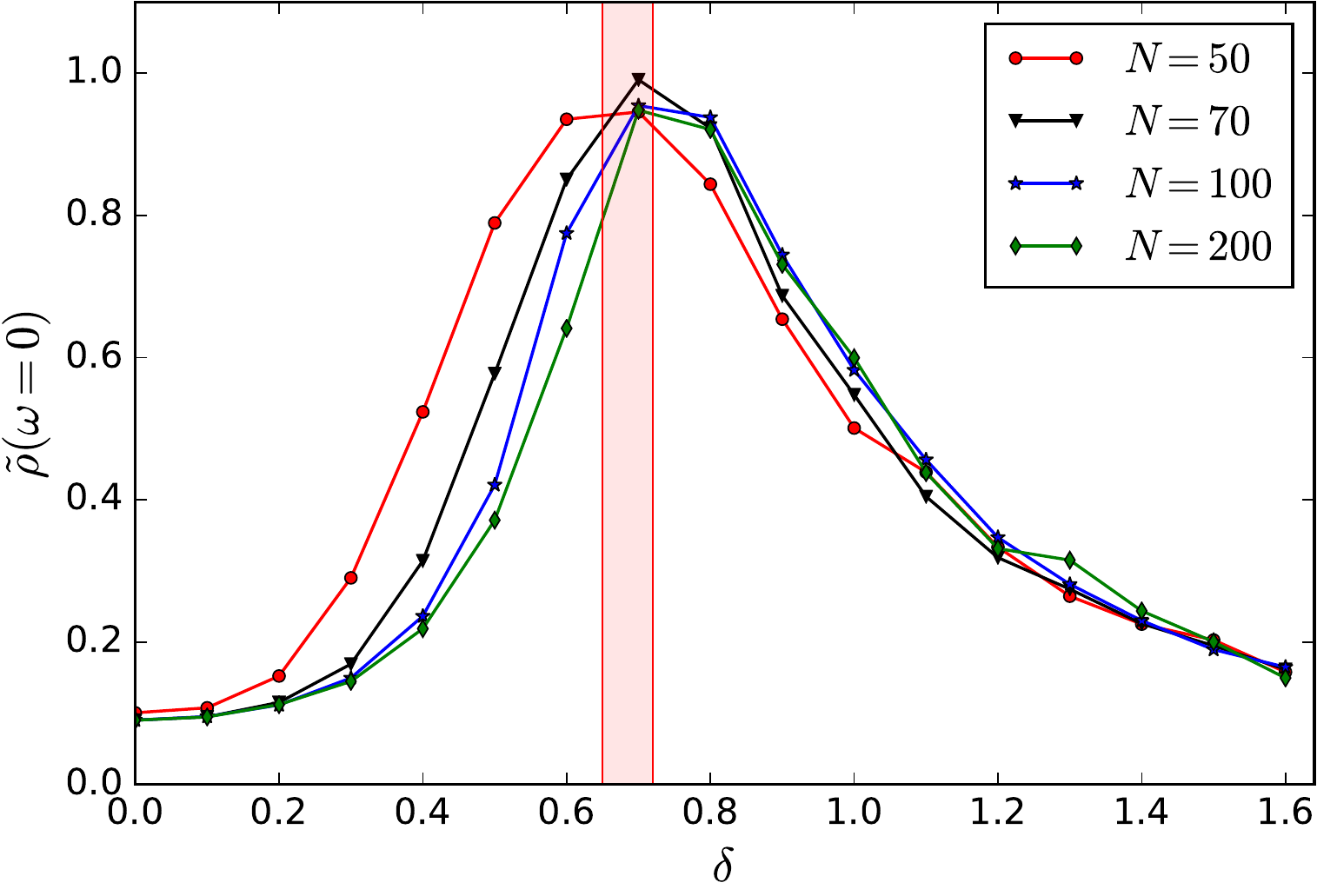}
\caption{Geometrically averaged local density of states 
obtained for $\omega=0$ at the central site $i=N/2$. 
The meaning of the red vertical strip 
is the same as in Fig.~\ref{middle_point}).}
\label{fig:tdos_srodek}
\end{figure}

Fig.\ \ref{fig:tdos_srodek} presents the geometrically averaged density 
of states at the central site $i=N/2$. We suspect that the weak disorder 
regime is related with the Majorana mode(s) which are eventually pinned 
at individual sites inside the chain. This process is somewhat sensitive 
to the atomic chain length $N$. On the other hand, for stronger disorder 
$\delta>\delta^{*}$, we  observe the monotonous decrease of the 
spectral function at zero energy (that is almost insensitive to $N$). 
Using  geometrical averaging we notice that the maximum 
of $\tilde{\rho}(\omega=0)$ is much more pronounced than 
in the case of arithmetic averaging (Fig. \ref{middle_point}). 
Moreover, in the present Fig. \ref{fig:tdos_srodek} the maximum 
perfectly coincides with the transition where the topological 
$Z_{2}$ number changes. It suggests that the local density of 
states in internal part of the Rashba chain (distant from the  
Majorana quasiparticles) could be useful for identifying  
the disorder--induced topological transition.

\section{Two--band Rashba chain\label{sec:two-band}}

\begin{figure}[H]
\includegraphics[width=\linewidth]{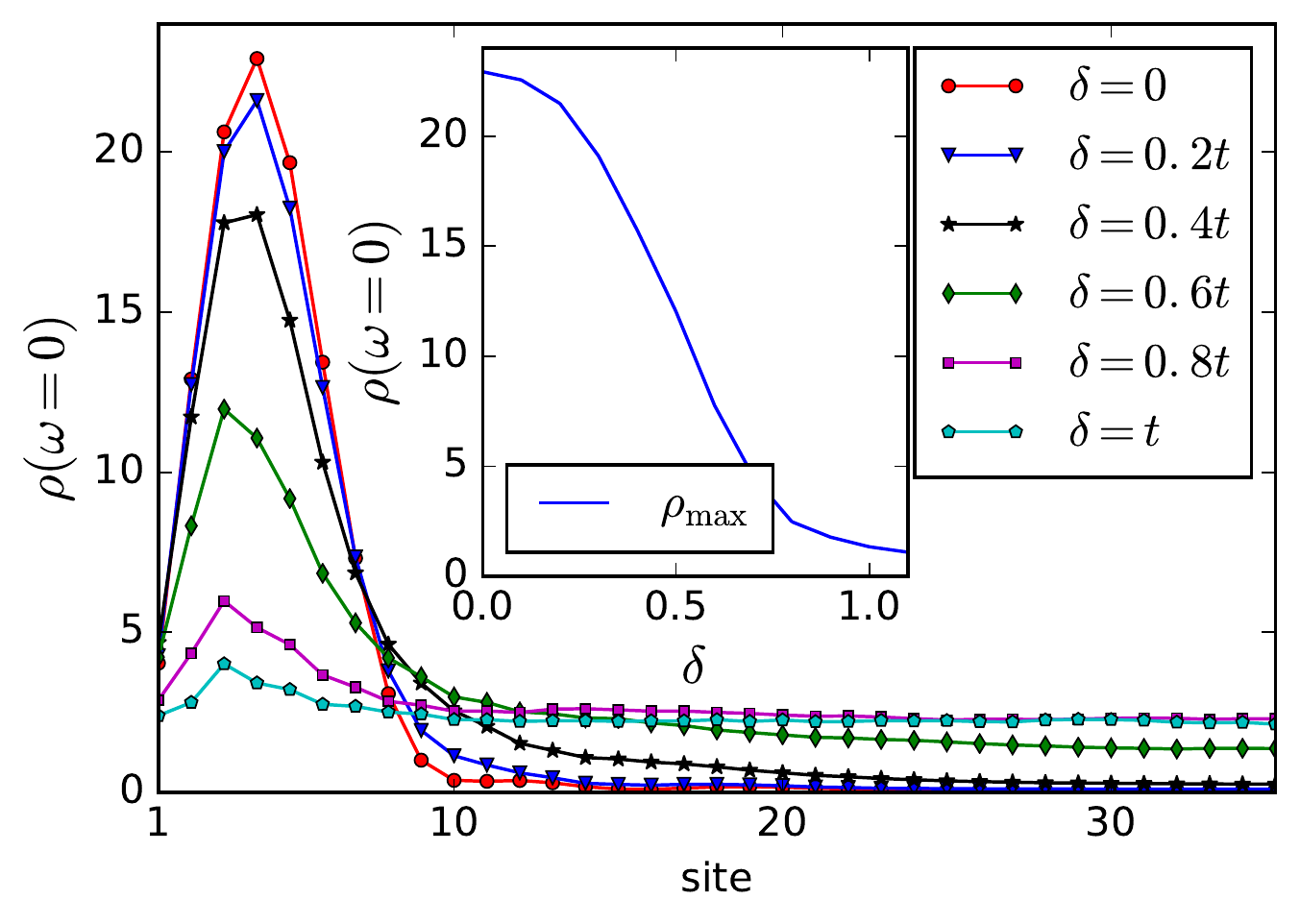}
\caption{The same as in Fig. \ref{diag_disorder}, but for a two--band model
  using $\mu = -1.05t$, $\mu_{\rm B}B=0.27t$, the interband hybridization $t_{12}=t$,
  and the interband spin--orbitcoupling $\alpha_{12}=\alpha$.
\label{fig:majorana_2bands}}
\end{figure}

\begin{figure}[H]
\includegraphics[width=0.85\linewidth]{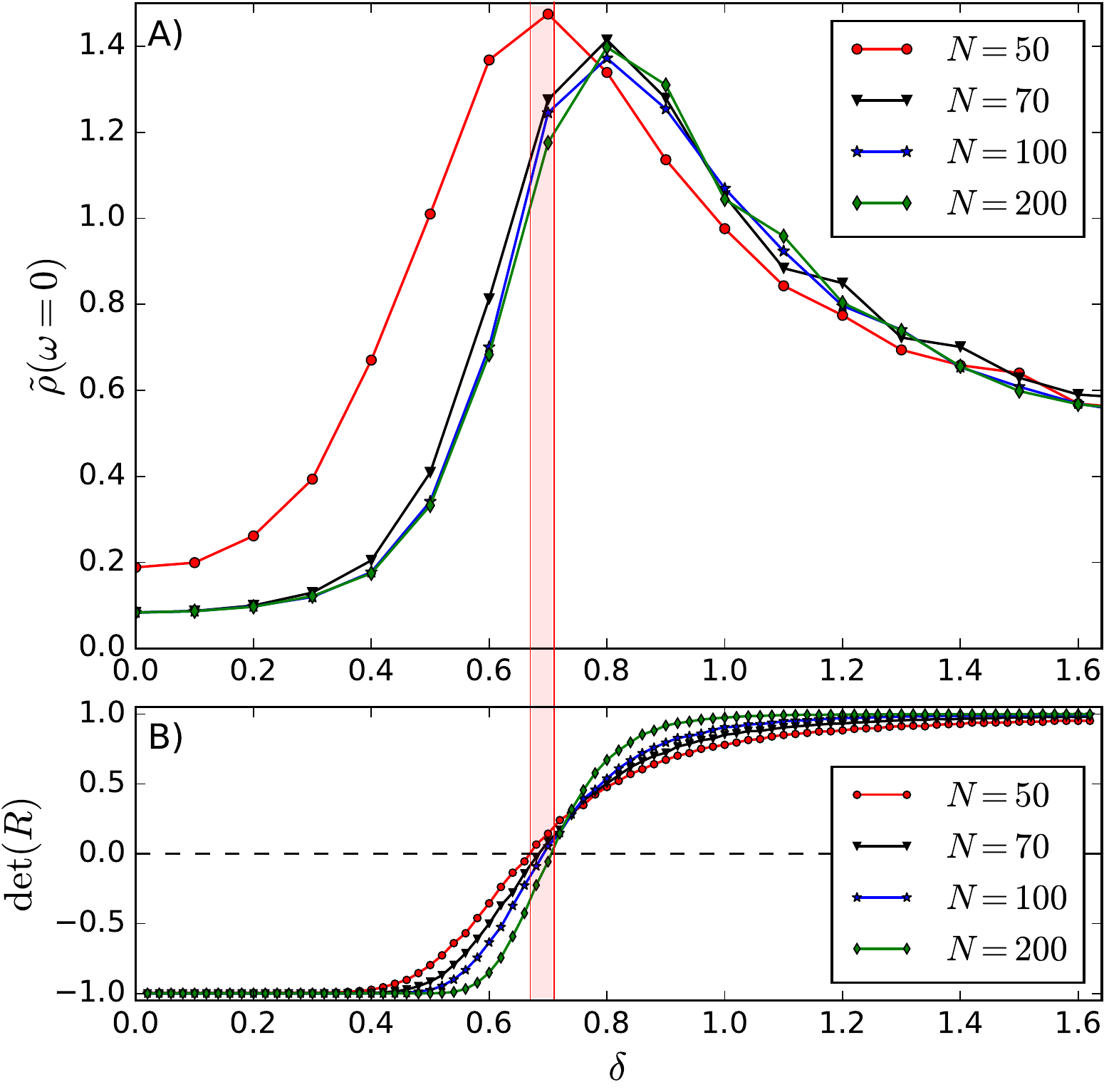}
\caption{Geometrically averaged local density of states 
at central site $i=N/2$ obtained for zero energy within 
the two--band model with the same model parameters as in Fig. \ref{fig:majorana_2bands}.
\label{fig:srodek_2bands}}
\end{figure}

Up to this point we investigated the Majorana quasiparticles of
a strictly one--dimensional chain, where they are created by
occupying only the lowest transverse sub--band of atomic nanowire.
Since such a condition might be difficult to satisfy in real experiments, 
we shall address here influence of disorder on the Majorana states 
for the simplest multi-band system. It has been demonstrated in 
Refs.\ \onlinecite{Lutchyn-2011a,Lutchyn-2011b} that a multiband 
Rashba chain coupled to $s$--wave bulk superconductor can be effectively 
modelled with two chains, whose coupling depends on external 
 parameters that can be tuned. For studying how robust  are 
 the Majorana quasiparticles of multiband disordered systems   
we generalize the Hamiltonian (\ref{proximized}) to the form 
\begin{eqnarray}
    \hat{H}^{\rm (prox)}_{\rm 2b}
    &=&\hat{H}_{\rm chain}^{\rm (prox)}(\hat{c}) 
    +\hat{H}_{\rm chain}^{\rm (prox)}(\hat{d}) 
    + t_{12}\sum_{i,\sigma}
    \hat{c}_{i\sigma}^\dagger\hat{d}_{i\sigma} + {\rm h.c.} \nonumber \\ 
    &+& \alpha_{12}\sum_{i,\sigma\sigma'}\left[\hat{c}_{i\sigma}^\dagger
    \left(i\sigma^x\right)\hat{d}_{i\sigma'}-\hat{d}_{i\sigma}^\dagger
    \left(i\sigma^x\right)\hat{c}_{i\sigma'}\right],
\end{eqnarray}
where $\hat{c}_{i\sigma}$ ($\hat{d}_{i\sigma}$) are annihilation operators 
for electrons in chain 1 (2) described by Hamiltonians $\hat{H}_{\rm chain}^{\rm (prox)}(\hat{c})$ [$\hat{H}_{\rm chain}^{\rm (prox)}(\hat{d})$]. 
The model parameters are tuned, to allow for emergence of the Majorana 
quasiparticle. Our results are presented in Figs. \ref{fig:majorana_2bands} 
and \ref{fig:srodek_2bands}. 

One can see, that dependence of the local density of states on 
disorder is almost exactly the same as in the one-band model. Again, 
the maximum of $\rho(\omega=0)$ in the central part of the chain   
coincides with the critical disorder $\delta^*$. This suggests 
that in multi-band case $\delta^*$ corresponds to transition 
between the topologically nontrivial and trivial phases.

\begin{figure}[H]
\vspace*{1mm}
\includegraphics[width=\linewidth]{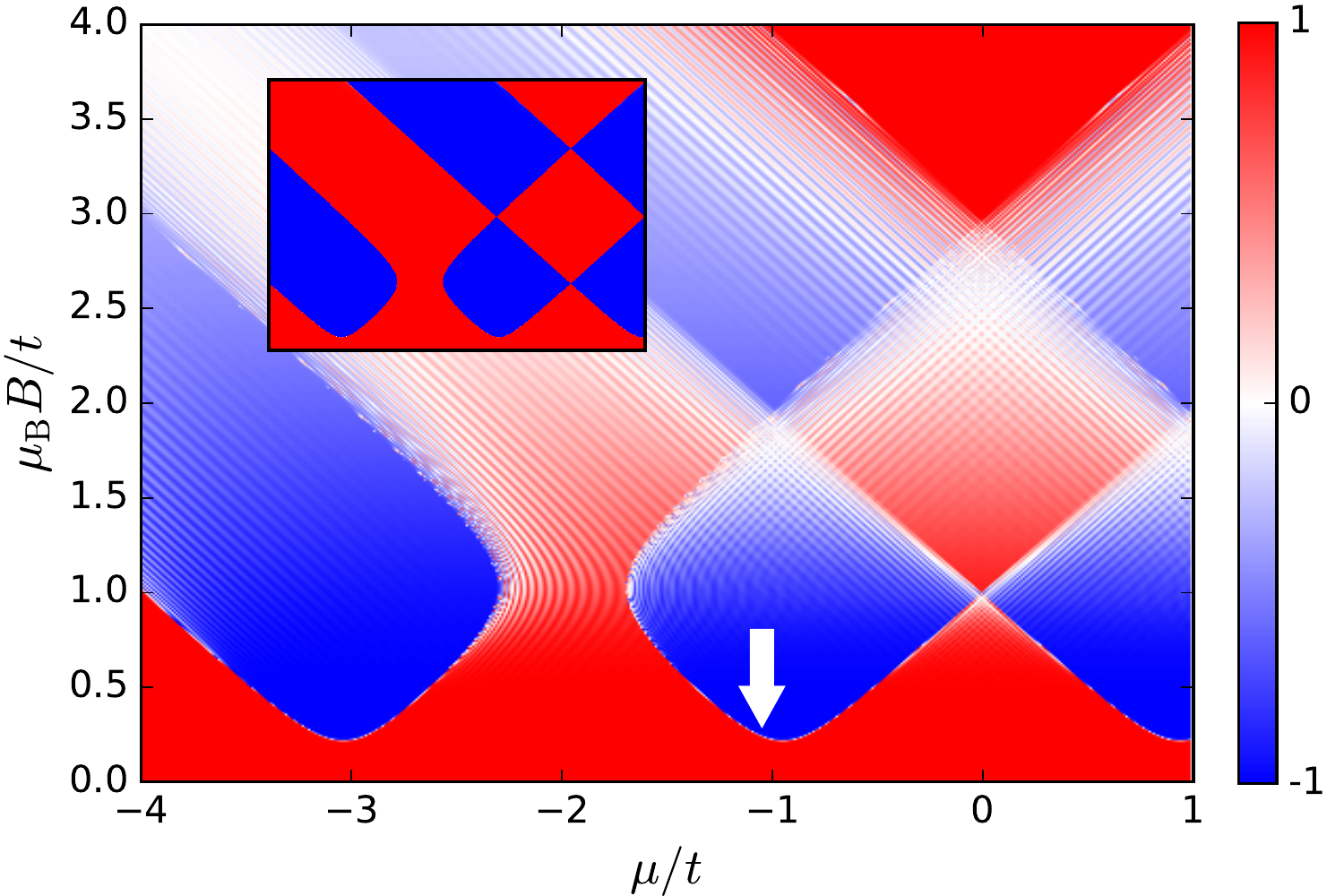}
\caption{The same as in Fig. \ref{fig:Q1}, but for a two--band 
model. The results presented in Figs.~\ref{fig:majorana_2bands} 
and \ref{fig:srodek_2bands} are obtained for the system  in 
topologically nontrivial regime (indicated by the white arrow), 
that is hardly affected by the finite size effects.
\label{fig:Q_2bands}}
\end{figure}

To confirm this observation we calculated the $\mathbb{Z}_2$ topological 
quantum number, following the procedure described in Sec.\ \ref{topo}. 
In the present case the scattering matrix [Eq. (\ref{matrix_s})] was 
composed of $8\times 8$ blocks $R,R',T,T'$ expressed in the basis given 
by vectors $\left(\hat{c}_{i\uparrow}^\dagger,\hat{c}_{i\downarrow}^\dagger, 
\hat{c}_{i\downarrow}, -\hat{c}_{i\uparrow},
\hat{d}_{i\uparrow}^\dagger,\hat{d}_{i\downarrow}^\dagger, 
\hat{d}_{i\downarrow}, -\hat{d}_{i\uparrow}\right)$. 
Numerical product of these matrices is even more unstable 
than for the one-band model case, but fortunately the method proposed 
in Ref. \onlinecite{Snyman-2008} helps to obtain reliable results.
Fig. \ref{fig:Q_2bands} shows $\det(R)$ for the uniform atomic chains 
comprising 70 and 500 lattice sites.

\begin{figure}[H]
\includegraphics[width=\linewidth]{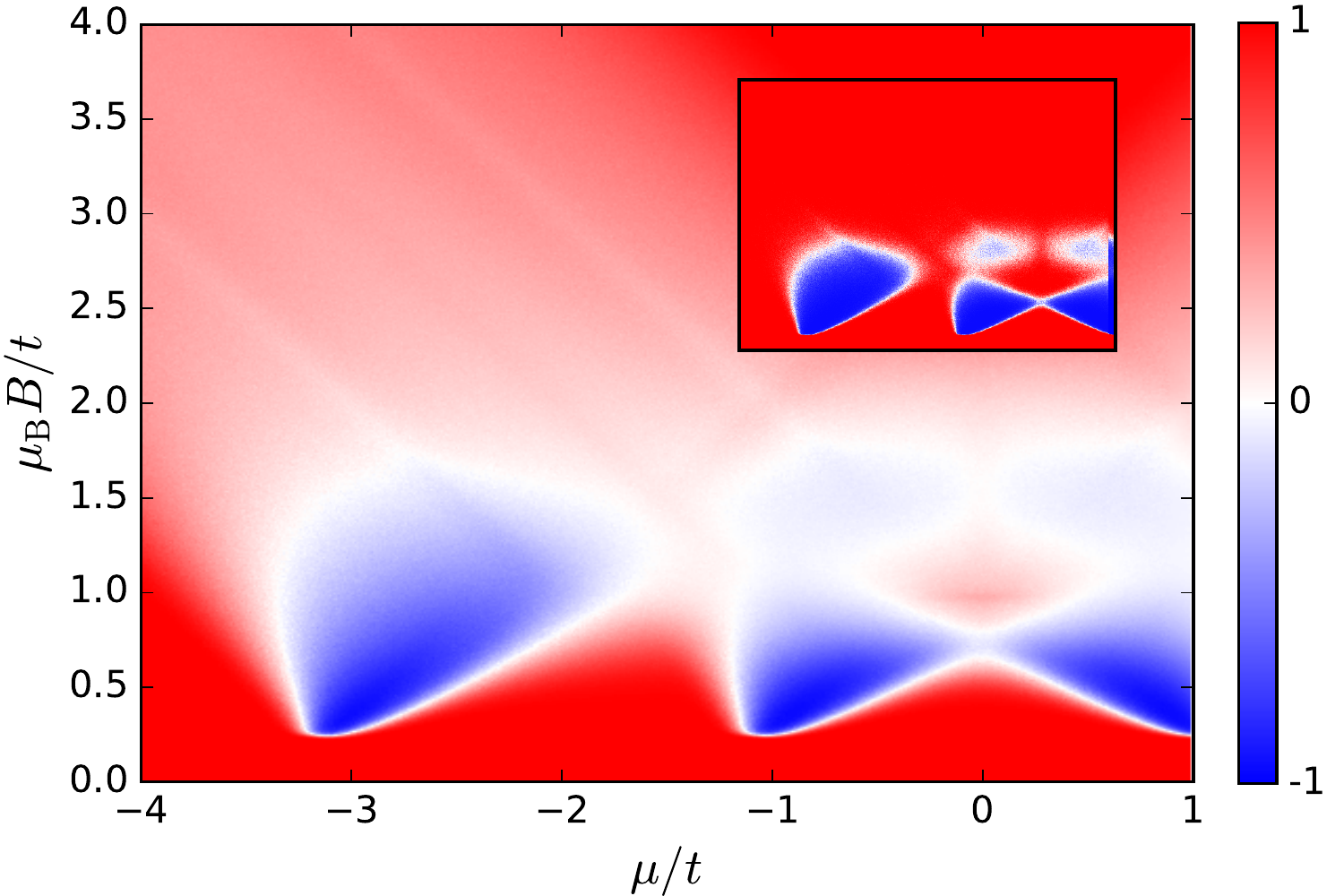}
\caption{The same as in Fig. \ref{fig:Q2}B, but for a two--band model. 
Additionally, the inset shows the sign of $\det(R)$ averaged over 
$10^4$ disorder realizations.
\label{fig:Q1_2bands}}
\end{figure}

Influence of the disorder is presented in 
Fig.~\ref{fig:Q1_2bands}. By comparing it with Fig. \ref{fig:Q2}B 
we notice, that the topologically nontrivial region is in both 
(one-- and two--band) models quite similar. The panels A and B 
in Fig. \ref{fig:srodek_2bands} indicate, that the disorder--induced 
topological transition coincides with the maximum of the local 
density of states (at zero--energy) in a central site of 
the atomic chain.

\end{document}